\documentclass[12pt,letterpaper]{article}

\usepackage[includeheadfoot,
            marginratio={1:1,2:3}, 
            width=412pt, 
            height=688pt,]{geometry}

\usepackage{amsmath}
\usepackage{amsfonts}
\usepackage{amssymb}
\usepackage{graphicx}
\usepackage{cite}

\newcommand{\eq}[1]{\begin{equation}
                     \begin{split} #1 \end{split}
                     \end{equation}}

\newcommand{\ov}{\overline}
\newcommand{\ostar}{\otimes_\star}
\newcommand{\tri}{\hspace{1.5pt}\triangle\hspace{1.5pt}}

\allowdisplaybreaks[2]
\numberwithin{equation}{section}

\begin{document}

\vspace*{-1.5cm}
\begin{flushright}
  {\small
  MPP-2016-50\\
  }
\end{flushright}

\vspace{1.5cm}

\begin{center}
{\LARGE
Towards a Theory of Nonassociative Gravity\\[0.3cm]
}
\vspace{0.4cm}

\end{center}

\vspace{0.35cm}
\begin{center}
  Ralph Blumenhagen and Michael Fuchs
\end{center}

\vspace{0.1cm}
\begin{center} 
\emph{Max-Planck-Institut f\"ur Physik (Werner-Heisenberg-Institut), \\ 
   F\"ohringer Ring 6,  80805 M\"unchen, Germany } \\[0.1cm] 

\vspace{0.4cm} 
 
\end{center} 

\vspace{1cm}


\begin{abstract}
Violating the strong constraint of double field theory, non-geometric
fluxes were argued to give rise to noncommutative/nonassociative
structures. We derive in a rather pedestrian physicist
way a differential geometry on the simplest nonassociative
(phase-)space arising for a constant non-geometric $R$-flux.
This provides  a complementary presentation to the quasi-Hopf
representation categorial one delivered by Barnes, Schenkel, Szabo in
arXiv:1409.6331+1507.02792. As there, the notions of tensors, covariant
derivative, torsion and curvature find a star-generalization.
We continue the construction  with the introduction of a
star-metric and its star-inverse
where, due to the nonassociativity, we encounter major deviations from
the familiar structure.  Comments on the Levi-Civita connection, a
star-Einstein-Hilbert action and the relation to string theory are
included, as well.
\end{abstract}


\clearpage
\tableofcontents

\section{Introduction}

For the open string it is a long established result that its effective
theory on a fluxed D-brane can be described by a noncommutative
gauge theory. Whereas, for the closed string the appearance
of a similar relation to noncommutative geometry is still under
debate. 
In \cite{Blumenhagen:2010hj,Lust:2010iy}, indications were presented
that support the picture that for
nongeometric closed string backgrounds the coordinates do not commute,
but give a noncommutative structure \cite{Lust:2010iy} for the case of so-called $Q$-flux
and a nonassociative structure \cite{Blumenhagen:2010hj,Lust:2010iy} 
for  the case of non-geometric $R$-flux
(see also the previous work \cite{Bouwknegt:2004ap} and
the reviews \cite{Plauschinn:2012kd,Blumenhagen:2014sba}).
This nonassociativity for a  constant $R$-flux background is captured by the
commutation relations
\eq{
                   [x^i,x^j]= \tfrac{i l_s^4 }{ 3 \hbar} \, R^{ijk} p_k\,,\qquad
                   [x^i,p_j]=i \hbar \, \delta^i{}_j \nonumber
}
where $p_k$ denotes the momentum and $i,j,k=1,\dots,D$. More evidence
for this result was delivered following
various alternative approaches \cite{Blumenhagen:2011ph,Condeescu:2012sp,Chatzistavrakidis:2012qj,Andriot:2012vb,Condeescu:2013yma,Bakas:2015gia}.

The string theoretic framework for nongeometric fluxes is double field
theory (DFT)
\cite{Siegel:1993xq,Siegel:1993th,Hull:2009mi,Hohm:2010jy}, 
an effective theory for the massless closed string modes
that features manifest $O(D,D)$ symmetry and is a priori defined
on a doubled space, where besides the usual coordinates $x^i$ one
introduces so-called winding coordinates $\tilde x_i$. The latter can be considered
as the canonical conjugate variables to the winding modes.  
For reviews of DFT please consult \cite{Aldazabal:2013sca,Berman:2013eva,Hohm:2013bwa}. 
As one of its peculiar features, DFT  is only consistent 
if one introduces a further constraint that 
reduces the degrees of freedom. For the fluctuations around
a given background, this has to be the strong-constraint
$\partial_i f \tilde \partial^i g+\tilde\partial^i f \partial_i g =0$
for every pair of fundamental objects $f$ and $g$.
This implies that eventually the quantities depend
on half of the coordinates.

It was pointed out in \cite{Blumenhagen:2013zpa} that  the
nonassociative algebra above,  presumes a violation of the 
strong constraint. In other words, one only has such a non-trivial
structure, if the strong constraint between the background $R^{ijk}$
and fluctuations around  it is violated. Since under these
circumstances the background and the fluctuations
are treated differently, this also presumes  a background dependent version
of DFT, similar to the one proposed in
\cite{Blumenhagen:2014gva,Blumenhagen:2015zma}. 
However, we emphasize
that the question about the correct form of the DFT constraints  is not completely settled
yet\footnote{An alternative proposal for a relation of DFT to a
  star-product
in noncommutative geometry was presented in \cite{Deser:2014wva,Deser:2015okd}.}. 

Once outside of Pandora's box, such a nonassociative structure has
gained some interest also from the more formal noncommutative geometry
point of view. In particular,  in the framework of
deformation quantization, a nonassociative star-product was
introduced that realizes the above algebra 
\cite{Mylonas:2012pg,Bakas:2013jwa,Mylonas:2013jha}. Since it originates from
the closed string, the expectation is that it might be possible to
define a gravity theory on such a background. Based on Hopf algebra
techniques, for the associative 
Moyal-Weyl star-product, a deformed formalism was 
developed \cite{Aschieri:2005yw,Aschieri:2005zs,AlvarezGaume:2006bn}
that  allows to generalize all the concepts from differential geometry, like
tensors, covariant derivative, torsion and curvature. The main idea
is to introduce so-called star-diffeomorphisms and deform
the Leibniz rule in such a way that the star-product of two star-tensors
is again  a star-tensor. 

Based on earlier work
\cite{Mylonas:2012pg,Bakas:2013jwa,Mylonas:2013jha}, this latter framework was generalized
recently by Barnes, Schenkel, Szabo \cite{Barnes:2015aa,Barnes:2015bb,Barnes:2016cjm}
to so-called 
quasi-Hopf algebras, which are not
any longer associative but whose associator is of a special form
that, as an example, includes the constant $R$-flux star-product.
The (to our taste) very  formal, categorial approach followed
in \cite{Barnes:2015aa,Barnes:2015bb,Barnes:2016cjm} led to the star-generalization of 
tensors, covariant derivative and curvature but stopped
at the point where usually a metric and its Levi-Civita connection
is introduced. Moreover, it is far from obvious whether all
these structures have anything to do with string theory or DFT,
respectively. If they do, then from string theory/DFT perspective
the following issue needs to be resolved:
\begin{itemize}
\item{At each order in $\alpha'$ the string effective action is
    manifestly diffeomorphism invariant. (Similarly, the DFT action is
    invariant under generalized diffeomorphisms.) Therefore, one
   needs to understand how star-diffeomorphisms are 
     related to these classical symmetries.}
\end{itemize}
One possible way to resolve it  may be  the following observation made in
\cite{Blumenhagen:2013zpa} and further exploited in
\cite{Aschieri:2015roa}. Since  string
theory is described on-shell by a two-dimensional conformal field, it was argued
that on-shell any sign of nonassociativity should better be
absent. That means that, using the equations of motion, the additional terms
in the action resulting from the star-product should be
total derivatives. Moreover, 
in \cite{Blumenhagen:2011ph} the CFT for a constant metric
with a constant $R$-flux was constructed up to linear order
in the $R$-flux.  Via computing correlation functions
of  tachyon vertex operators a linear contribution was found
that could be encoded in the  nonassociative star-product from above.  
This  means that comparing the two formalisms is expected to be
reliable only up to linear order in the $R$-flux.

In this paper, even though we will not be able to fully clarify  the above
mentioned issue,
as an intermediate step, we intend to provide a more pedestrian derivation 
of the structure of a nonassociative differential geometry. 
Avoiding abstract techniques from quasi-Hopf algebras, we investigate  how a
star-tensor calculus can be developed step by step. We will work
on the entire phase space and will be able to construct a covariant
derivative, the torsion and the curvature star-tensor.
It is remarkable that this is still possible, even though,
due to the nonassociativity, one has to be very careful  with
the bracketing in all of the expressions that appear
in the course of the formal computations.
However, when it comes to the introduction of a star-metric,
its inverse and a Levi-Civita connection, we encounter a number of
obstacles  in  the nonassociative
case that were not present in the only noncommutative but associative
case. 

Concretely, this paper is organized as follows: In section 2 we introduce
the nonassociative star-product for a non-vanishing $R$-flux
and, after  discussing some of its basis structures,
we carry out  the first steps towards a star-tensor formalism.
Here we restrict to tensors only depending on configuration space.
In section 3 we generalize some of these concepts to a full 
phase-space dependence, before in section 4 we develop the
full tensor calculus. Section 5 is devoted to formulate the basic
notions of a nonassociative differential geometry, i.e.
we introduce a covariant derivative and define its torsion and
curvature. 
In section 6 we move on and introduce a star-metric and
discuss the appearing deviations from the usual structure
once one wants to introduce  a Levi-Civita connection and an
Einstein-Hilbert action.


\section{Nonassociative star-product}

In this  section  we provide some  basic
definitions and features of the nonassociative star-product and
introduce the concept of star-diffeomorphisms. 
The nonassociative star-product is not completely generic
in the sense that it does admit two important operators that
control the way in which the product is noncommutative and
nonassociative, respectively.
In this first warm-up section, for simplicity we restrict ourselves
to star-diffeomorphisms in configuration space $M$, i.e. those
without an explicit momentum dependence. The generalization
to the full phase space ${\cal M}=T^*M$ will be presented in section 3.
As mentioned in the introduction,  we will refrain from
using the rather abstract Hopf algebra techniques from
\cite{Mylonas:2012pg,Bakas:2013jwa,Mylonas:2013jha, Aschieri:2015roa, Barnes:2015aa, Barnes:2015bb,Barnes:2016cjm},
but try to build up the formalism step by step in an explicit way.
To complete the picture, a brief introduction into quasi-Hopf algebras
is presented in appendix \ref{app_hopf}.

\subsection{The universal ${\cal R}$-matrix and the associator $\phi$}

In \cite{Mylonas:2012pg}, a star-product was introduced that upon deformation
quantization leads to the nonassociative algebra presented
in the introduction. It can be considered as a nonassociative 
generalization of the Moyal-Weyl star-product.
Thus, throughout this paper we will work with the  star-product
\begin{eqnarray} 
\label{starproduct}
&&\hspace{-0.7cm} f \star g := \cdot \left[ {\cal F}^{-1} (f, g) \right]\\
&&\quad= \cdot \Big[ \exp \left( \tfrac{1}{2} i \hbar ( \partial_i \otimes \tilde \partial_p^i -  \tilde \partial_p^i \otimes \partial_i)  + \tfrac{i l_s^4}{12 \hbar} R^{ijk} \big( p_k \partial_i \otimes \partial_j - \partial_j \otimes p_k \partial_i  )\right) \,  f \otimes g \Big] \nonumber
\end{eqnarray} 
which was  suggested to capture the presence of a totally antisymmetric
$R^{ijk}$-flux background in string theory. 
Note that in contrast to Moyal-Weyl, this star-product lives in the
full phase space ${\cal M}$ containing  derivatives in the space  directions
$\partial_i$  and the momentum directions $\tilde \partial_p^i$.
We used the tensor product to indicate  on which factor of $f\otimes g$
the derivatives act with  the dot in front eventually turning  the tensor products
into usual multiplications. 
More explicitly, the star-product reads
\eq{
f \star g &= f \cdot g + \tfrac{1}{2} i \hbar  ( \partial_i f\, \tilde \partial_p^i g-  \tilde \partial_p^i f\,  \partial_i g ) + \tfrac{i l_s^4}{6 \hbar} R^{ijk} p_k\, \partial_i f \,\partial_j g + \dots \,\, .
}
The first part of the product leads to the Heisenberg commutation relations when inserting $f = x^i$ and $g = p_i$ and the second part gives the non-trivial commutator between the coordinates
\eq{
[x^i, x^j ] = \tfrac{i l_s^4 }{ 3 \hbar} R^{ijk}\,p_k .
}
This product is nonassociative and violates the Jacobi identity. 
It was proposed in \cite{Mylonas:2013jha} that, when higher tensors are multiplied,
the partial derivatives have to be replaced by Lie derivatives to guarantee compatibility with the exterior derivative. With
$f,g$ now higher tensors, the star-product is defined as
\eq{ \label{starwithlie}
f \star g = f \cdot g &+ \tfrac{1}{2} i \hbar  ( {\cal L}_{\partial_i} f\, {\cal L}_{\tilde \partial_p^i } g + {\cal L}_{\tilde \partial_p^i} f \, {\cal L}_{\partial_i} g )\\&+ \tfrac{1}{2} \big(   \tfrac{i l_s^4}{6 \hbar} R^{ijk}\, {\cal L}_{ p_k\partial_i} f \,{\cal L}_{\partial_j} g - \tfrac{i l_s^4}{6 \hbar} R^{ijk}\, {\cal L}_{\partial_j} f \,{\cal L}_{p_k \partial_i} g  \big) \dots \,\, .
}
The non-trivial part is the ${\cal L}_{ p_k\partial_i} f$ in the last
line. 

The operator ${\cal F}^{-1}$ we defined in \eqref{starproduct} is called the twist. As it is given by a phase its inverse ${\cal F}$ can be read off by switching the sign in the exponent. Due to the antisymmetry in the exponent a permutation of the arguments also inverts the twist
\eq{
{\cal F}^{-1}(f, g) = {\cal F} (g, f).
}
Using this we can deduce  what happens to the star product under a permutation
\eq{
\label{rmatrixdef}
f \star g =& \cdot \big[ {\cal F}^{-1} (f, g) \big] = \cdot \big[ {\cal F} (g, f) \big] = \cdot \big[ {\cal F}^{-1} \underbrace{{\cal F} {\cal F}}_{:= \ov{\cal R}} (g, f) \big]\\:=& \ov{\cal R}(g) \star \ov {\cal R}(f) .
}
Here we introduced the universal ${\cal R}$-matrix ${\cal R} = {\cal
  F}^{-2}$ whose inverse $\ov {\cal R}  = {\cal F}^{2}$ captures the
extra factors appearing when one permutes scalars in 
the star-product. Therefore, it is  a representation
of the permutation group on this algebra. The  notation in the second
line of \eqref{rmatrixdef} means
that first the $\ov {\cal R}$-matrix acts on $g$ and $f$ and
afterwards the star-product is carried out. 
To see what is happening here, we evaluate \eqref{rmatrixdef} for functions
$f,g\in C^\infty(M)$ on configuration space and up to linear order
in the $R$-flux. For $f\star g$ we have
\eq{
     f\star g=f g+ \tfrac{i l_s^4}{6 \hbar} R^{ijk} p_k\, \partial_i f
     \,\partial_j g + \dots
}
and for $\ov{\cal R}(g) \star \ov {\cal R}(f) $
\eq{ \label{concreteRaction}
    \ov{\cal R}(g) \star \ov {\cal R}(f) &=
     g\star f-2  \tfrac{i l_s^4}{6 \hbar} R^{ijk} p_k\, \partial_i g \star
     \,\partial_j f + \dots\\
    &=g f + \tfrac{i l_s^4}{6 \hbar} R^{ijk} p_k\, \partial_i g
     \,\partial_j f -2  \tfrac{i l_s^4}{6 \hbar} R^{ijk} p_k\, \partial_i g 
     \,\partial_j f + \dots\\
   &=fg + \tfrac{i l_s^4}{6 \hbar} R^{ijk} p_k\, \partial_i f
     \,\partial_j g + \dots\,.
}

The second important object is the associator $\phi$ which reorders the brackets in a product of three functions
\eq{ \label{associator}
(f \star g) \star h = f^{\phi} \star (g^{\phi} \star h^{\phi}) := f \star (g \star h) |_{\phi}.
}
The inverse associator $\ov \phi$ similarly shifts brackets to the left.
Again the associator is acting on the functions first and then the star product is executed. The associator is central and commutes with ${\cal F}$ and ${\cal R}$. Therefore expressions like ${\cal R}(f)^{\phi} = {\cal R}(f^{\phi}) $ are unambiguous. Explicit calculation gives the phase factor
\eq{ \label{defassociator}
\phi(f, g, h) = \exp \Big( \tfrac{l_s^4}{6} R^{ijk} {\cal L}_{\partial_i} \otimes {\cal L}_{\partial_j} \otimes {\cal L}_{\partial_k} \Big) (f \otimes g \otimes h).
} 
Thus \eqref{associator}  can be expressed in more detail as
\eq{
   (f \star g) \star h = f \star (g \star h) +\sum_{n=1}^\infty
   \frac{1}{n!} (\tfrac{l_s^4}{6})^n\, &R^{i_1 j_1 k_1}\ldots R^{i_n j_n k_n}
                   (\partial_{i_1}\ldots \partial_{i_n} f)\star \\ 
               &\Big(
                 (\partial_{j_1}\ldots \partial_{j_n} g)\star
                  (\partial_{k_1}\ldots \partial_{k_n} h)\Big)\,.
}
For instance, for three functions $f,g,h\in C^\infty(M)$ on
configuration space,  we find the following  terms up to second order in $R^{ijk}$
\eq{
\label{hertha}
(f \star g) \star h =& f \star (g \star h)  \\& + \tfrac{l_s^4}{6} R^{ijk} \, \partial_i f \;\; \partial_j g \;\; \partial_k h \\& + \tfrac{l_s^4}{6} \tfrac{i l_s^4}{6 \hbar} R^{ijk} R^{abc} \, p_c \, (\partial_i f \;\;  \partial_a \partial_j g \;\; \partial_b \partial_k h + \text{cycl.} )\\
&+ {\cal O} (R^3)\,.
}
The inverse associator $\ov \phi$ is obtained by switching the sign in the exponent. Due to the antisymmetry of  $R^{ijk}$ also a permutation of the arguments inverts the twist. In this way the ${\cal R}$-matrix can invert the associator in our notation
\eq{\label{RinverstsAssociator}
f^{\phi} \star (g^{\phi} \star h^{\phi} ) = f^{\ov \phi} \star ( \ov{\cal R}(h^{\ov \phi}) \star \ov{\cal R}(g^{\ov \phi}) ) 
}
as we have $\phi(f, g, h) = \ov \phi(f, h, g)$.

Our nonassociative star-product is special in the sense that it admits the two operations
${\cal R}$ and $\phi$ that capture the effect of noncommutativity and
nonassociativity. As we will see in section \ref{algebra1}, precisely these two extra
operations allow one to still  write down a generalized Leibniz rule 
for star-diffeomorphisms.

\subsection{Derived tri-products}
\label{sec_tri}

Clearly, in contrast to the star-product on the Moyal-Weyl plane, the
most  unconventional aspect of this star-product is that
it involves the momentum coordinates. In \cite{Blumenhagen:2011ph}, the CFT
for a constant $R$-flux on a flat space allowed to extract
information on a non-trivial tri-product on configuration space.
The relation between the star- and the tri-product was 
suggested in \cite{Aschieri:2015roa}.

Indeed, using the star-product \eqref{starproduct}, one can define so-called tri-products for
functions on configuration space  via
\eq{
\label{triproda}
           f_1 \tri f_2 &\tri \ldots \tri f_N =
            f_1\star (f_2\star ( \ldots (f_{N-1}\star
            f_N)\ldots ))\big\vert_{p_0=0}\\
      &=\cdot\Big[ \exp\Big( -\tfrac{l_s^4}{12} \sum_{1\le a<b<c\le
            N} R^{ijk} \,\partial^a_i 
             \otimes \partial^b_j \otimes \partial^c_k \Big)
           (f_1\otimes \ldots \otimes f_N)\Big]\,.
}
Note that after evaluating the nested star-product,  one is
restricting the result to the $p_0=0$ leaf. This implies e.g.
$f_1\tri f_2=f_1\, f_2$.
These tri-products have the peculiar property that all $R$-flux
dependent corrections  become total derivatives so that the integral 
drastically simplifies as
\eq{
\label{triprodb}
                \int d^d x\, f_1 \tri f_2 \tri \ldots
                \tri f_N = \int d^d x\, f_1\, f_2\, \ldots f_N\,.
}
As proposed in \cite{Blumenhagen:2013zpa}, this fact can provide a way that
nonassociativity of the underlying space-time can trivialize in the
action. 
In \cite{Aschieri:2015roa} they also argue
 that string theory might be  realized on the $p_0 = 0$ leaf, while
 deviations from $ p_0 = 0$ correspond to membrane corrections.

\subsection{Scalars and the Leibniz Rule}
\label{algebra1}

Following \cite{Aschieri:2005yw,Aschieri:2005zs,AlvarezGaume:2006bn},
in this section  we take the first steps towards developing a
star-tensor formalism. 
One defines a star-scalar as  an object that transforms
under a star-diffeomorphism, $\xi=\xi^i(x) \star \partial_i$  with the star-Lie derivative
\eq{
           \delta_\xi f= {\cal L}^\star_\xi f:=\xi^i\star \partial_i f\,.
}
We will consider only diffeomorphisms in the space directions first
and discuss momentum diffeomorphisms with a full $\xi = \xi^i
\star \partial_i + \tilde \xi_i \star  \tilde \partial_p^i $
later. Next one demands that  the star-product of two
star-scalars should again be a star-scalar,
i.e. we enforce 
\eq{
\label{prodtensr}
      \delta_\xi (f\star g)= {\cal L}^\star_\xi (f\star
      g)=\xi^i\star \partial_i ( f\star g)=\xi^i\star (\partial_i
      f\star g)+
     \xi^i\star  ( f\star \partial_i g) \, . 
}
Here we used that the star product adds factors of the momentum but
not of the coordinates such that the derivative can be pulled
through. We can use the ${\cal R}$-matrix and the associator to bring this into the form of a generalized Leibniz rule. In the first term an inverse associator is enough while in the second term we find
\eq{
\xi^i\star  ( f\star \partial_i g) &= (\xi^i\star   f)\star \partial_i g |_{\ov \phi} \\
&= (\ov{\cal R}(f) \star \ov{\cal R}(\xi^i))\star \partial_i g |_{ \phi} =  \ov{\cal R}(f) \star (\ov{\cal R}(\xi^i)\star \partial_i g) |_{ \phi^2}. 
}
From the first to the second line the permutation of the arguments inverts the associator similar to \eqref{RinverstsAssociator}. From this we recognize 
\eq{
\label{leibnizr} 
    {\cal L}_\xi^\star (f\star
      g)=  \big({\cal L}^\star_{\xi^{\ov\phi}} f^{\ov\phi}\big) \star g^{\ov\phi}+
     \ov{\cal R}(f^{ \phi^2}) \star \big( {\cal L}^\star_{\ov{\cal R}(\xi^{ \phi^2})} g^{\phi^2} \big).
}
Now we can proceed by defining higher star-tensors.

\subsubsection*{Vectors and covectors}

Next, we require that $\omega_i=\partial_i f$ is a star-covector.
The behavior under star-diffeomorphisms can be deduced as
\eq{
    \delta_\xi (\partial_i f)=\partial_i (\delta_\xi   f)=\xi^j\star (\partial_j\partial_i f )+ (\partial_i
    \xi^j) \star (\partial_j f)
} 
so that a covector $\omega_i$ generically transforms as
 \eq{
\label{covectorbe}
          \delta_\xi \omega_i ={\cal L}^\star_\xi \omega_i &=\xi^j\star (\partial_j \omega_i )+ (\partial_i
    \xi^j) \star \omega_j\,.
}
Following the general logic, a star-vector $v^i$ is defined via
\eq{
      \delta_\xi v^i ={\cal L}^\star_\xi v^i &=\xi^j \star (\partial_j v^i )- \ov{\cal R}(v^j) \star \ov{\cal R}(\partial_j
    \xi^i) \,.
}
This guarantees that $f=v^i \star \omega_i $ transforms as a star-scalar.
This gives rise to the definition of the star-commutator of star-vectors
\eq{ \label{commutator}
                   [v,w]_\star=v^j\star (\partial_j w^i )-
                \ov{\cal R} (w^j) \star   \ov{\cal
                  R}(\partial_j v^i) 
}
which is equal to  the Lie-derivative.
Clearly the commutator is manifestly $\ov{\cal R}$-antisymmetric $[v,w]_\star=-[\ov{\cal R}( v),\ov{\cal R} (w)]_\star$. This commutator could also be defined by $[\,,]_\star := [\,,] \circ {\cal F}^{-1}$ making its covariance obvious through the similarity to the definition of the star-product.
\subsubsection*{Tensors}
Next we define the star-tensor-product of e.g. two star-vectors
as $z=u \ostar v$ with $z^{ij}=v^i \star w^j$.
Extending the Leibniz rule \eqref{leibnizr} to the tensor product,
\eq{
\label{leibniztensorr}
    {\cal L}^\star_\xi (v\ostar
      w)=  \Big({\cal L}_{\xi^{ \ov\phi}}^{\star} v^{\ov \phi}\Big) \ostar w^{\ov\phi}+
     \ov{\cal R}(u^{\phi^2}) \ostar \Big( {\cal L}^{\star }_{\ov{\cal R} (\xi^{\phi^2})} w^{\phi^2} \Big)
}
one finds
\eq{
{\cal L}_\xi^\star z^{ij} &=\xi^k \star (\partial_k z^{ij} )- (\partial_k
    \xi^i) \star z^{k j}- (\partial_k
    \xi^j) \star z^{i k}\,.
}
The generalization to higher tensors is straightforward.

\subsubsection*{Composition and closure}

When two Lie-derivatives act on an object we face a bracketing
ambiguity.  The correct solution lies in the commutator of the
Lie-derivative. The closure property 
\eq{
             [{\cal L}^\star_\xi,{\cal L}^\star_\eta]_\star\, v=
                                {\cal L}^\star_ {[\xi,\eta]_\star}\, v\,
}
is fulfilled if we define the commutator of two star-Lie-derivatives as
\eq{ \label{commutatorLieder}
        \left [{\cal L}^\star_\xi,{\cal L}^\star_\eta \right]_\star\, v:= {\cal
          L}^\star_{\xi^{\phi}} \big(
          {\cal L}^\star_{\eta^{\phi}}
         \, v^{\phi}\big)-
        {\cal L}^{\star}_{\ov{\cal R}(\eta^{ \phi})}
        \big({\cal L}^{\star}_{\ov{\cal R}(\xi^{ \phi})}
         \, v^{ \phi}\big)\,.
}
This matches a known Hopf algebra result. There the consistent composition $\bullet$ of operators has the crucial property
\eq{ \label{composition}
(O \bullet O')(z) = O^{\phi} \big( O'^{\phi} (z^{\phi})\big)\,.
}
This rule is very intuitive as the associator reorders the brackets in the usual way. This composition operator obviously enters the commutator of the Lie-derivatives and as we will see later also in the Riemann tensor. Thus the commutator is in general given by
\eq{
[A,B] = A\bullet B - \ov{\cal R}(B) \bullet \ov{\cal R}(A)\,.
}

\subsection{Comment on star-scalars}
\label{fundamentalfunctions}

Since we have defined the Leibniz rules such that the star-product
of two star-scalars is again a star-scalar, the question occurs
whether an ordinary function $f(x)\in C^\infty(M)$ on configuration
space is actually a star-scalar. 
To approach this question, let us consider the Taylor expansion of
such a function
\eq{
                   f(x)=\sum_{n_1,\ldots,n_k=0}^\infty    a_{n_1,\ldots,n_k}   \,       x_1^{n_1}\cdot \ldots \cdot x_k^{n_k}\,.
}
Taking into account that $x_i\star x_i=x_i \cdot x_i$, we can define a
corresponding star-Taylor expansion as
\eq{
\label{startaylor}
                   f_\star(x)=\sum_{n_1,\ldots,n_k=0}^\infty
                   a_{n_1,\ldots,n_k}   \,       x_1^{n_1}
              \star ( x_2^{n_2} \star (\ldots \star   x_k^{n_k})\ldots )\,.
}
Since the elementary, linear functions $h(x)=x_i$ are star-scalars
by construction $f_\star(x)$ in \eqref{startaylor} is a star-scalar, as well.
Now we can define a subset of all star-scalars as ${\mathfrak F}=\{
f: f_*=f\}$, i.e. the set of those star-scalars,
where the star-multiplications in the Taylor expansion act trivially.

Let us show by induction that $\exp(q_i x^i)\in {\mathfrak F}$ with
$q_i={\rm const}$.  Clearly, the linear function $g(x)=\vec q\cdot
\vec x\in {\mathfrak F}$. Let us assume that $h_n(x)=(\vec q\cdot
\vec x)^n\in {\mathfrak F}$.  We need to show that $g(x)\star h_n(x)=g(x)
h_n(x)$. Since $g(x)$ is linear, the star-product simplifies to
\eq{
    g(x)\star h_n(x)&=g(x) h_n(x)+R^{ijk}\, \partial_i g(x)\, \partial_j
    h_n(x)\, p_k\\
   &=g(x) h_n(x)+\left( R^{ijk}\, q_i q_j p_k\right)  g(x)\, n h_{n-1}(x)\\
   &=g(x) h_n(x)\,.
}
Therefore, the tachyon vertex operators $\exp(\vec q \,\vec x)$, from
which features of the star-product were derived in \cite{Blumenhagen:2011ph}, 
are indeed star-scalars. This can be considered as a  
self-consistency check. Now it is clear that every ``sum'' of such terms
\eq{
             f(x)=\int d^k q  \hat f(q)\, e^{q_i x^i}
}
is also a star-scalar. Therefore, every  $f(x)\in C^\infty(M)$ is a
star-scalar as $f(x)\in {\mathfrak F}$.

\section{The  star-product on phase space}

Since in the star-product \eqref{starproduct} both the space coordinates and the
momenta appear, it actually is defined on the phase space ${\cal M}$.
Therefore the restriction to configuration space from the previous
section is not very natural.
Moreover, also vectors should have components along the
full tangent space, including the momentum directions.
As a splitting into space and momentum components results in a lot of
terms,  we introduce  a doubled notation where $X_I=(\tfrac{p_i}{i \hbar},x^i)$ and
\eq{
V = V^I(X) \star \partial_I = V^i(x,p) \star \partial_i + \tilde V_i(x,p) \star i\hbar\, \tilde \partial^i _p\,.
}
Similar to double field theory, a sum over a capital index
$I=1,\ldots,2D$ always runs over an  upper and a lower index.
For revealing its full contents we will present most formulas in
both a doubled and in a split-into-components notation.
In this section we will derive  several useful formulas that will be
used in the next section for the construction of a nonassociative
differential geometry calculus.

\subsection{Action on the basis}

Recall that in the star-product \eqref{starwithlie} several Lie-derivaties appear which usually reduce to partial derivaties for functions. More concretely we have
\eq{ \label{operators}
{\cal L}_{\partial_i} :=& P_i \,,\\
i \hbar \, {\cal L}_{\tilde \partial_p^i} :=& \tilde P^i \,,\\
 \tfrac{i l_s^4 }{ 6  \hbar} R^{ijk} \,{\cal L}_{{p_j} \partial_k} :=& M^i\, .
}
$P$ and $\tilde P$ denote translations in space and momentum
directions,  while $M$ induces so-called Bobb-shifts in the momentum directions. Using these operators we can rewrite the star-product as
\eq{ \label{starwithoperators}
f \star g
&= \cdot \left[ \exp \big( \tfrac{1}{2}  ( P_\mu \otimes \tilde  P^\mu -  \tilde P^\mu \otimes P_\mu)  +  \tfrac{1}{2} \big( M^\mu \otimes P_\mu - P_\mu \otimes M^\mu ) \big) f \otimes g \right]  . 
}
In doubled notation we can merge the operators from \eqref{operators} into
\eq{
P^I &= (\tilde P^i, P_i) = 
( i \hbar \,\,\tilde {\cal L}_{\partial_p^i}\,\,,\,\, {\cal L}_{\partial_i})  \,,\\
M^I &= - F^{IJK} X_J P_K =  ( \tfrac{i l_s^4}{6\hbar} {\cal L}_{R^{ijk} p_j \partial_k }, 0)\,, 
}
where $F^{IJK}$ has only one non-vanishing component $F^{ijk} =  \tfrac{l_s^4 }{6} R^{ijk}$. 
The fact that these operators contain Lie-derivatives implies a
non-trivial action on the fundamental forms $dx^i$, $dp_i$.
Merging them   into $dX^I = ( dx^i, \tfrac{dp_i}{i \hbar})$
and  using $[{\cal L}, d] = 0$, one obtains
\eq{ \label{actiononbasis2}
P^I (dX^J) &= 0\,,\\
M^I (dX^J) &= F^{IJK} dX_K \qquad \text{or} \qquad 
M^i (dx^j) =  \tfrac{l_s^4}{6} R^{ijk} \, \tfrac{dp_k}{i\hbar}\, .
}
By duality between tangent and cotangent space or by acting on 
$\delta_J{}^I = \partial_J \star dx^I
= \partial_j \star  \, dx^i + i \hbar \, \partial_p^j \star \tfrac{dp_i}{i \hbar}$ one
has the relations
\eq{ \label{actiononbasis}
P^I (\partial^J) &=  0\,,\\
M^I (\partial^J) &=   F^{IJK} \partial_K \qquad \text{or} \qquad M^i(i \hbar \, \tilde \partial^j_p) = \tfrac{l_s^4}{6 } R^{ijk} \partial_k\,. 
}
Note that $e^I=\partial^I$ is considered here as a basis of the tangent space.
Let us make the following three observation:
\begin{itemize}
\item{Since the $P_i$ act trivially onto any basis vector, the
    associator becomes the identity when acting on a basis vector. We will refrain from writing brackets in an expression like $A \star B \star \partial_I$ when both bracketings are equal.}
\item{However, the $M^i$ and therefore the $\star$-product itself and the
    ${\cal R}$-matrix act non-trivially on the basis vector
    $\tilde \partial_p$.}
\item{Due to
    $M(M(\tilde \partial_p)) = M(\partial) = 0$ 
       the star-product terminates after the first order when acting
    on any basis vector.}
\end{itemize}
The main task in the following  is to take this non-trivial action
of $M^i$ into account when developing the star-tensor calculus and
a star-differential geometry.
Since there are only first order corrections, this is still 
feasible.

\subsection{Star-commuting scalars and vectors with basis vectors}

Let us now consider expressions like $f(X)\star \partial_I$.
For any arbitrary function $f= f(x,p)$ we trivially find\footnote{Note
  that, in
  contrast to $\partial_i f$, the derivative  in $\partial_i \star f=e_i \star f$
 is not meant   to act on $f$.}
\eq{ \label{switching2}
f \star \partial_i = f \cdot \partial_i   \quad \Rightarrow \quad f \star \partial_i - \partial_i \star f &=0 \,,
}
but an additional term arises when expanding the star-product \eqref{starwithoperators} in 
\eq{ \label{switching}
f \star i \hbar \, \tilde \partial_p^i = f \cdot i \hbar \, \tilde\partial_p^i - \tfrac{1}{2} P_j f \cdot M^j(i \hbar \, \tilde \partial_p^i) = f \cdot i \hbar \, \tilde\partial_p^i   + \tfrac{l_s^4}{12} R^{ijk} \partial_j f \star \partial_k \,.
}
so that
\eq{
 f \star i \hbar \, \tilde \partial^i_p - i \hbar \, \tilde \partial^i_p \star f &=  \tfrac{l_s^4}{6 } R^{ijk} \partial_j f \star \partial_k \,.
}
In doubled notation this can be written as
\eq{
\label{hannover96}
f \star \partial^I - \partial^I \star f &= F^{IJK} \partial_J f \star \partial_K\,. 
}
By comparing \eqref{actiononbasis2} with \eqref{actiononbasis} we see
that the star-product acts in the same way on $dx^I$ and
$\partial^I$. The relation \eqref{hannover96} must therefore also hold for the basis forms
\eq{\label{switching3}
f \star dx^I - dx^I \star f &= F^{IJK} \partial_J f \star dx_K\,.
}
From \eqref{hannover96} we see that the
two possibilities to form a vector, $V^I \star \partial_I$ and
$\partial_I \star V^I$, are not equivalent. As will be explained later
we  use the first convention. 
This has  also consequences for the right and left 
multiplication of a scalar and a vector. While for the left
multiplication
one gets 
\eq{
f \star (V^I \star  \partial_I) = (f \star V^I) \star \partial_I\,,
}
since the associator vanishes on basis vectors but 
the right multiplication gives an additional term 
\eq{
 ( V^I \star \partial_I) \star f = (V^I \star f) \star  \partial_I 
 - F^{IJK}\, \,  (V_I \star \partial_J f) \star \partial_K\,.
}

The non-trivial action onto the basis vector  also has consequences
for  the derivation of  the Leibniz rule. Recall that in \eqref{prodtensr}
we were exchanging the order of a basis vector $e_i=\partial_i$ 
and a scalar. 
Therefore, using \eqref{hannover96} we can derive the rule
\eq{
\label{werder}
\partial^I (f \star g) &= \partial^I f \star g + f \star \partial^I g - F^{IJK} \partial_J f \star \partial_K g \,.
}
The same formulas continue to  hold more generally for the operators
$P^I$ where the partial derivatives are replaced by Lie-derivatives. 
The commutator with the Bopp-shifts  is simply
\eq{
[P^I, M^J] &= F^{IJK} \,  P_K\,,
}
reproducing  the action on the basis vectors \eqref{actiononbasis}. 
With this commutator at hand one can calculate the Leibniz rule for $M^I$
\eq{ 
M^I (f \star g) = M^I f \star g + f \star M^I g  + F^{IJK} \, \,  \partial_J f \star \partial_K g \,.
}

Finally, the non-trivial star-commutators between the basis vectors
and the scalars or vectors have non-trivial consequences when 
an expression involves  the action of the ${\cal R}$-matrix.
Looking back one realizes  that the structure of the additional
terms arising
from the action on the basis vectors  can essentially be deduced
immediately from the index  structure.
For the ${\cal R}$ matrix acting on a scalar and a vector, one can
derive the relation
\eq{ \label{Vector}
\ov{\cal R}(f) \otimes \ov{\cal R}(V^I \star \partial_I) = \ov{\cal R} (f) \otimes \ov{\cal R}(V^I) \star \partial_I + F^{IJK} \, \,  \ov {\cal R}(\partial_I f) \otimes \ov {\cal R}(V_J) \star \partial_K\,,
}
and by iteration we find for the interchange of two vectors
\eq{ \label{Ronvectors}
\ov {\cal R}(V^I \star \partial_I ) \otimes \ov {\cal R}(U^J \star \partial_J)&= 
\ov {\cal R}(V^I) \star \partial_I \otimes \ov {\cal R}(U^J) \star \partial_J \\
&- F^{MNI}\, \,  \ov {\cal R} (V_N) \star \partial_I \otimes \ov {\cal R} (\partial_M U^J) \star \partial_J \\
&+ F^{MNJ} \, \,  \ov {\cal R} (\partial_M V^I) \star \partial_I \otimes    \ov {\cal R}  (U_N)\star \partial_J  \\
&-  F^{ABI} F^{MNJ} \, \,  {\cal R}(\partial_M V_B) \star \partial_I \otimes   {\cal R}(\partial_A  U_N) \star \partial_J\,.
}

As we will see in section 4, even though  this non-trivial action
onto the basis vectors  will produce a lot of additional terms, 
they will nevertheless organize themselves perfectly well
so that the form of the star-tensor relations from section 2 remain
the same \footnote{Such a nice behavior  was shown 
in \cite{Barnes:2015aa,Barnes:2015bb} for
arbitrary quasi-Hopf algebras.}.

\subsection{Star-pairing between vectors and forms}

Let us move on and define a star-pairing  between star-vectors and
star-forms.  Eventually, this should result in a contraction of the
components with a star-product  in between, i.e.
\eq{ 
\label{contration}
V^I \star \omega_I = V^i \star \omega_i + \tilde V_i \star \tilde \omega^i\,.
}
First, let us define the product between the basis vectors and forms in the easiest way.
\eq{ 
\label{contractionbasis}
\partial_I \star dx^J &= dx^J \star \partial_I = \delta_I^J \,.
}
In order to obtain  the intuitive contraction \eqref{contration} as a
consequence of this basis  vector multiplication, we need the
following convention for forms  and vectors. In star-vectors the basis
vectors must be on the right and in forms they must be on the left 
\eq{
V &= V^I \star \partial_I \,,\qquad
\omega = dx^I \star \omega_I  \,.
}
Only then we indeed find
\eq{
V \star \omega = (V^I \star \partial_I) \star (dx^J \star \omega_J) = V^I \star \delta_I{}^J \star \omega_J &= V^I \star \omega_I \,.
}
This convention for forms and vectors corresponds to the mathematical definition of the contraction
\eq{ \label{pairing}
\langle\,, \rangle_\star : TM \otimes_\star T^*M &\rightarrow {\mathbb
  R}, \quad  \\ V \otimes_\star \omega &\rightarrow \langle\,, \rangle
\circ {\cal F}^{-1} (V, \omega) = V^I \star \omega_I 
}
where the first entry is reserved for the vector and the second one for the form. 

\section{The star-tensor calculus}

In this and the next section  we will develop the basic notions of a nonassociative
differential geometry on the full twisted phase space. From a
more mathematical  perspective this was already done in
\cite{Barnes:2015bb} 
for arbitrary quasi-Hopf algebras. The purpose here is  to show rather
explicitly  how this concretely works for the star-product \eqref{starproduct}.
We first repeat the procedure from section
\ref{algebra1} and develop the notion of star-diffeomorphisms and its 
corresponding tensors. Here we need to rely on the relations derived in
the previous  section.

\subsection{Scalars}

We define a star-scalar $f = f(x,p)$ to be a quantity transforming
under twisted diffeomorphisms generated by a vector $\xi = \xi^I(X) \star \partial_I = \xi^i (x,p) \star \partial_i + \tilde \xi_i (x,p)  \star i\hbar \, \tilde \partial_p^i$ as
\eq{
\delta_\xi f = {\cal L}^\star_\xi f = \xi^I \star \partial_I f = \xi^i \star \partial_i f + \tilde \xi_i \star i \hbar \, \tilde \partial_p^i f\,.
}
This definition is the reason for our choice of convention $V^I \star \partial_I$ instead of $\partial_I \star V^I$.
We demand the star-product of two scalars to be a star-scalar again. Switching the partial derivatives through with \eqref{werder} gives
\eq{ \label{productscalars}
{\cal L}^\star_\xi (f \star g) &=\xi^I \star \partial_I (f \star g) \\
       &=
             \xi^I \star (\partial_I f \star g)+\xi^I \star
             (f\star \partial_I  g) - F^{IJK} \xi_I \star (\partial_J f
             \star \partial_K g)\,.
}
Utilizing the relation  \eqref{Vector},  this expression including the
$F^{IJK}$-correction term can be compactly written as
\eq{
{\cal L}^\star_\xi (f \star g) =& \left( {\cal L}^\star_\xi f\right)
\star g \big\vert_{\ov \phi} + \ov {\cal R}(f) \star \left( {\cal L}^\star_{\ov {\cal R} (\xi)} g\right)\big\vert_{\phi^2}\,.
}
The Lie-derivative ${\cal L}^*$ therefore still obeys the same
Leibniz rule as in \eqref{leibnizr}. The extra terms arising from the
non-trivial
commutators with the basis vectors conspire in such a way that the
formal expression for the Leibniz rule does not change.
This is quite a remarkable feature.

\subsection{Vectors and covectors}

Guided by \eqref{commutator} we define the transformation of a vector
$V$ with the twisted commutator
\eq{
\delta_\xi V = {\cal L}^\star _\xi V = [\xi, V]_\star  := [\,,] \circ {\cal F}^{-1}\,.
}
This definition guarantees a covariant behavior under star multiplications and reads
\eq{\label{trafovector}
{\cal L}^\star _\xi V & = \xi ( V) - \ov {\cal R}(V) \big(\ov{\cal R}(\xi)\big)\\ &= \xi^I \star \partial_I V^J \star \partial_J - \ov{\cal R}(V)^I \star \partial_I (\ov {\cal R}(\xi)^J) \star \partial_J\, .
}
In \eqref{Ronvectors} we computed how the ${\cal R}$-matrix acts on two vectors. Inserting this in the definition of the commutator \eqref{trafovector} gives in components
\eq{ \label{trafovectorcomp}
{\cal L}^\star _\xi V^I= &
\xi^J \star \partial_J V^I-\ov {\cal R}(V^J) \star \partial_J \big(  \ov {\cal R}(\xi^I) \big) \\[0.1cm]
&-  \ov {\cal R} ( \partial_M V^J) \star \partial_J \big(  \ov {\cal R}  (\xi_N) \big)  F^{MNI} \,.
}
Here, we did not pull the ${\cal R}$-matrices out of the momentum
derivatives as more terms proportional to ${ F}^{IJK}$ would
arise. 

Next we verify the closure property of two
star-Lie-derivatives acting on a star-scalar
\eq{ \label{closure}
[{\cal L}_\xi^\star, {\cal L}_\eta^\star]_\star f = \xi(\eta(f)) |_{\phi} - \ov{\cal R}(\eta) \big( \ov {\cal R}(\xi)(f) \big)|_{\phi} = {\cal L}^\star_{[\xi, \eta]_\star}\,.
}
For this purpose we compute
\eq{ \label{vectorsonf}
(\xi \bullet \eta) f &= \xi^{\phi} ( \eta^{\phi} f^{\phi}) = \xi^I \star \partial_I ( \eta^J \star \partial_J f) |_\phi \\[0.2cm]
&= (\xi^I \star \eta^J) \star \partial_I \partial_J f + (\xi^I
\star \partial_I \eta ^J) \star \partial_J f \\
&\phantom{=}\ - F^{IJK} (\xi_I \star \partial_J \eta^A )\star \partial_K \partial_A f
}
and 
\eq{
(\ov{\cal R} \eta \bullet \ov{\cal R} \xi) f 
=& (\xi^J \star \eta^I) \star \partial_J \partial_I f + (\ov{\cal R}(\eta^I) \star \partial_I \ov{\cal R}(\xi^J) ) \star \partial_J f  \\
& +  ( F^{MNJ} \ov {\cal R} (\partial_M \eta^I) \star      \partial_I \ov {\cal R}  (\xi_N) ) \star \partial_J f \\
& +  (F^{MNJ}    \xi_N \star   (\partial_M \eta^I)) \star  \partial_I \partial_J f \,.
}
Adding both terms,  remarkable cancellations occur that
finally yield  \eqref{closure}. 
Although intuitive  it is tedious to verify the Leibniz rule
\eq{
[U, [V, W]_\star ]_\star = [[U, V]_\star, W]_\star |_{\ov \phi} + [\ov {\cal R}(V), [\ov{\cal R}(U), W]_\star]_\star |_{\phi^2}
}
that is nothing else than the star-Jacobi identify for three star-vectors.
This relation can be found more easily with Hopf algebra techniques as
in \cite{Aschieri:2015roa}.

Next we come to the definition of a star-covector $\omega=dx^I
\star\omega_I$. Like in section 2, its star-Lie-derivative can be deduced from the variation of the partial derivative of a star-scalar. Before doing so we want to introduce a derivative operator $\partial$ that will become very handy. Recall the product rule for the partial derivative \eqref{werder}  
\eq{
\partial_I (f \star g) &= \partial_I f \star g + f \star \partial_I g - F_I{}^{JK} \partial_J f \star \partial_K g \,.
}
Multiplying this relation by $dx^I$ and interchanging $dx^I$ with $f$
using \eqref{switching3}, one obtains
\eq{ \label{leibnizforpartial}
dx^I \star \partial_I (f \star g) &= dx^I \star \partial_I f \star g +
dx^I \star f \star \partial_I g 
 - F^{IJK} dx_I \star  \partial_J f \star \partial_K g \\
&= dx^I \star \partial_I f \star g + f \star dx^I \star \partial_I g - F^{IJK} dx_K \star \partial_J f    \star \partial_I g \\
& \hspace{5.58cm} - F^{IJK} dx_I \star  \partial_J f \star \partial_K g \\
&= dx^I \star \partial_I f \star g + f \star dx^I \star \partial_I g\,.
} 
We observe that the correction terms have canceled and we are left with the usual Leibniz rule for the combination $dx^I \star \partial_I$. It is therefore suggestive to introduce the derivative operator
\eq{ \label{defderivative}
\partial := dx^I \star \partial_I\,.
}
$\partial$ raises the degree by one but since we do not assume antisymmetrization $\partial$ is not the exterior derivative $d$.
Using $\partial$, \eqref{leibnizforpartial} turns into the usual Leibniz rule
\eq{ \label{extdercommute}
\partial(f \star g) = \partial f \star g + f \star \partial g\,.
}
Therefore $\partial$ commutes with the star-product and also the ${\cal R}$-matrix. This will simplify many computations from now on.

Let us come back to the variation of the partial derivative of a scalar. Having $\partial$ at hand, we instead compute the variation of the one-form $\partial f = dx^I \star \partial_I f$. Only the use of $\partial$ allows us to interchange $\delta^\star_\xi$ with $\partial$ without producing extra terms. We get
\eq{
\delta^\star_\xi \partial f &= \partial \delta^\star_\xi  f \\
&=  \partial(\xi^I \star \partial_I f) = \partial \xi^I \star \partial_I f + \xi^I \star \partial_I \partial f\,,
}
from which  we deduce for a form $\omega = dx^I \star \omega_I$
\eq{
{\cal L}^*_\xi \omega &= \xi^I \star \partial_I \omega + \partial \xi^J \star \omega_J \,
}
or in components
\eq{ \label{formstransform}
{\cal L}^*_\xi \omega_I =&  \xi^J \star \partial_J \omega_I + \partial_I \xi^J \star \omega_J\,.
}
We explicitly checked that the star-Lie-derivatives of vectors  \eqref{trafovectorcomp} and forms \eqref{formstransform} are compatible with the contraction. Therefore a contraction between a vector $V$ and a form $\omega$ indeed transforms as a scalar when computing
\eq{
{\cal L}^*_\xi(V^I \star \omega_I) = {\cal L}^*_\xi V^I \star \omega_I |_{\ov \phi} + \ov{\cal R} (V^I) \star {\cal L}^*_{\ov {\cal R}(\xi)} \omega_I |_{\phi^2}\,.
}

\subsection{Tensor product}

Let us now define and investigate the notion of a $\star$-tensor
product in more detail.
In order for tensor products to behave covariantly under star-diffeomorphisms we define $\otimes_\star := \otimes \circ {\cal F}^{-1}$ similar to the definition of the commutator and the star-product. Then the Leibniz rule holds
\eq{
{\cal L}^*_{\xi}(V \otimes_\star W) = {\cal L}^*_\xi V \otimes_\star W|_{\ov \phi}  + \ov{\cal R}(V) \otimes_\star {\cal L}^*_{\ov{\cal R}(\xi)} W |_{\phi^2}\,.
}
To apply this rule to an arbitrary element for instance $T \in TM \otimes_\star TM$ one might need a split 
\eq{
T = A{}_{\alpha}{} \otimes_\star B{}^\alpha \in TM \otimes_\star TM
}
with an internal summation over $\alpha$ and $A_{\alpha} =
A_{\alpha}^I \star \partial_I\in TM$ and $B^{\alpha} =
B^{\alpha I} \star \partial_I \in TM$. The transpose of the above $T$ is defined as $T^T = \ov{\cal R}(B^\alpha) \otimes_\star \ov{\cal R}(A_\alpha)$. When shifting the basis vectors of $A$ and $B$ to the right with \eqref{hannover96}
\eq{ \label{basistotheright}
T &= A_\alpha ^I \star \partial_I \otimes_\star B^{\alpha J} \star \partial_J \\ &= (A_\alpha^I \star B^{\alpha J} - F^{MNJ} A_{\alpha M} \star \partial_N B^{\alpha I} ) \star (\partial_I \otimes_\star \partial_J )
}
and using \eqref{Ronvectors}, we find that the transpose $T^T$ only interchanges the indices due to nice cancellations 
\eq{  \label{Rbasistotheright}
T^T &= \ov{\cal R}(B^{\alpha I} \star \partial_I) \otimes_\star \ov{\cal R}(A^J_\alpha \star \partial_J)\\ &= (A_\alpha^J \star B^{\alpha I} - F^{MNI} A_{\alpha M} \star \partial_N B^{\alpha J} ) \star (\partial_I \otimes_\star \partial_J ) \, . 
}
Therefore ${\cal R}$-symmetric and ${\cal R}$-antisymmetric tensors still correspond to symmetric and antisymmetric matrices.

The contraction of  tensor products is done by multiplying the forms and vectors standing next to each other with \eqref{contractionbasis}. Most easily this can be done by bringing the basis vectors and forms to the middle as in \eqref{basistotheright} and eventually applying 
\eq{ \label{higherscalarpr}
\langle \partial_I \otimes_\star \partial_J, dx^A \otimes_\star dx^B
\rangle_\star =  \delta_I ^B \, \delta_J ^A\,.
}
For $A \otimes_\star B \in TM^2$ and $\omega \otimes_\star \alpha \in T^*M^2$ 
we get
\begin{eqnarray}
\label{scalarbetweentensors}
\langle A\otimes_\star B \; ,\; \omega \otimes_\star \alpha
\rangle_\star\!\!\! &=&\!\!\!(A^I \star B^J) \star (\omega_J \star \alpha_I) -
F^{IJK} (A_I \star \partial_J B^M) \star (\omega_M \star \alpha_K ) \nonumber\\[0.1cm]
&&\qquad - F^{IJK} (A_I \star B^M) \star ( \partial_J \omega_M \star \alpha_K ) \, . 
\end{eqnarray}
With \eqref{basistotheright} and \eqref{Rbasistotheright} and their analogue for forms one can show the transposition symmetry
\eq{ \label{symmetry}
\langle A\otimes_\star B \; ,\; \omega \otimes_\star \alpha \rangle_\star = \langle \ov{\cal R}(B) \otimes_\star \ov{\cal R}(A)\; ,\;  \ov{\cal R}(\alpha) \otimes_\star \ov{\cal R}(\omega) \rangle_\star\,.
}
The generalization to higher rank tensors is straightforward.

Antisymmetric $p$-forms $\omega \in \wedge_\star^p T^*M$ are  defined
as usual. Of course, here  one  requires  ${\cal R}$-antisymmetry and  adjusts
the star-wedge product to $\wedge_\star = \wedge \circ
{\cal F}^{-1}$. 
Considering,  for instance,  the star-wedge product of two 
one-forms one finds
\eq{ \label{wedgeantisymmetric}
\omega \wedge_\star \alpha = \omega \otimes_\star \alpha - \ov{\cal R}(\alpha) \otimes_\star \ov{\cal R}(\omega) 
}
which is clearly ${\cal R}$-antisymmetric.
From \eqref{symmetry} we see that the star-wedge product projects out the antisymmetric part
\eq{\label{twoformonvectors}
\langle A \otimes_\star B, \omega \wedge_\star \alpha \rangle_\star &= 
\langle A \otimes_\star B, \omega \otimes_\star \alpha \rangle_\star -
\langle \ov {\cal R}(B) \otimes_\star \ov{\cal R}(A), \omega \otimes_\star \alpha \rangle_\star
} 
so that one could  also  define forms by  the antisymmetry of their
action onto vectors. The exterior derivative is the antisymmetrized
partial derivative $d = \partial^{\wedge_\star} = dx^I
\wedge_\star \partial_I$. Inherited from $\partial$ and
\eqref{extdercommute} the exterior derivative $d$ is invariant under
the ${\cal R}$-matrix or 
\eq{
d(\omega \otimes_\star \alpha) = d \omega \otimes_\star \alpha + \omega \otimes_\star d \alpha\,.
}
This can also be explained from the fact that the star-exterior derivative is
the usual exterior derivative, as $d = dx^I \wedge_\star \partial_I =
dx^I \wedge \partial_I$. Since the star-product acts with usual
Lie-derivatives ${\cal L}$ satisfying  $[{\cal L}, d] = 0$, the exterior
derivative commutes with the star-product and therefore also with
${\cal R}$. This was already observed in \cite{Aschieri:2005zs}.

\subsection{Comment on $O(D,D)$ metric}

Since we are  using a doubled formalism with doubled
coordinates $X_I=(p_i,x_i)$ and doubled vector fields $V^I=(V^i,\tilde
V_i)$,
one might wonder what the relation to DFT is.
The main difference is that we are not dealing here with an $O(D,D)$ 
covariant formalism. This is reflected in the fact that instead of
generalized diffeomorphisms like in DFT we are dealing with only double
dimensional ordinary (star-)diffeomorphisms. 
One could be tempted to define an $O(D,D)$ metric between two vectors
via
\eq{
                      V^I \star W_I=V^i \star \tilde W_i +\tilde V_i\star W^i\,.
}
However, this is not a (star-)scalar under (star-)diffeomorphisms.

\section{Nonassociative differential geometry}

In this section we continue developing the basic notions of 
a nonassociative differential geometry. We will discuss  star-connections, its torsion and curvature tensors.

\subsection{Covariant derivatives}

The next step is to define a covariant derivative.
As we will see, in the nonassociative case there exist two consistent
notions of a covariant derivative, where  one acts from the left and
the other one from the right. 

First, we compute the anomalous variation $\Delta^\star _\xi :=
\delta_\xi^\star - {\cal L}^\star _\xi$ of the derivative $\partial
\omega$ of a star-covector\footnote{Here, $\partial$ means the one-form
  defined in \eqref{defderivative}.}
 \eq{ 
\label{variationdform}
\Delta^\star _\xi \; \partial\omega =  \partial \partial \xi^J \star  \omega_J\,.
} 
As usual we introduce a Christoffel-symbol $\Gamma$ which we can always write as 
\eq{ \label{Gammadef}
\Gamma = dx^I \star dx^J \star \Gamma_{IJ}{}^K \star \partial_K
}
by commuting the basis forms and vectors through with
\eqref{hannover96} and \eqref{switching3}. 
Now we can form the operator $\nabla = \partial -
\Gamma$ where $\Gamma$ acts with a star-contraction. Dealing with a
noncommutative star-product we can either let $\nabla$ act from the
left or from the right to form the covariant derivative. Due to
\eqref{extdercommute} this corresponds to the ambiguity from which
side we want to 
multiply $\Gamma$ 
\eq{
\overrightarrow{\nabla}  \omega &= \nabla (\omega) =  \partial \omega - \Gamma \star \omega \\
\text{or} \qquad \overleftarrow{\nabla}  \omega &= (\omega) \nabla = \partial \omega - \omega \star \Gamma\,.
}
For $\overrightarrow{\nabla}$ the star-contraction in the second term
is especially simple giving $ \Gamma^K \star \omega_K$ while for
$\overleftarrow{\nabla}$ we need an ${\cal R}$-matrix. Moreover the 
right-linearity of $\overrightarrow{\nabla}$ is reminiscent of the right-linearity
of $\omega= dx^I \star \omega_I$. Nonetheless both covariant
derivatives are completely  consistent.
Computing the anomalous variations $\Delta^\star _\xi :=
\delta_\xi^\star - {\cal L}^\star _\xi$ of the second terms, we find
that both choices correctly compensate the anomalous term  in
\eqref{variationdform} 
if $\Delta^\star_\xi \Gamma = \partial \partial \xi$ 
\eq{ \label{anomalousvariation}
\Delta^\star _\xi (\Gamma \star \omega) &= \Delta_\xi^\star \Gamma \star \omega = \partial \partial \xi \star \omega \,,\\[0.2cm]
\Delta_\xi (\omega \star \Gamma) &= \ov{\cal R}(\omega) \star
\Delta_{\ov {\cal R}(\xi)} \Gamma =  \ov{\cal R}(\omega)
\star \partial \partial \ov {\cal R}(\xi) \\
&= \ov {\cal R}(\omega) \star \ov{\cal R}( \partial \partial \xi ) = \partial \partial \xi \star \omega \,.
}
In the second line we used that the ${\cal R}$-matrix commutes with
$\partial$ according to \eqref{extdercommute}. 

Let us mention that
from a more axiomatic  viewpoint both choices are meaningful. Indeed, in
mathematics one defines the covariant derivative as a map $T^*M
\rightarrow T^*M \otimes_\star T^*M $ 
obeying the Leibniz rule
\eq{
\nabla(\omega  \otimes_\star \alpha) = \nabla \omega \otimes_\star \alpha |_{\ov \phi} + \ov{\cal R}(\omega) \star \ov{\cal R}(\nabla) \alpha |_{\phi^2}\,.
}
Taking into account that $\nabla f=\partial f$   for
scalars, for $\overrightarrow{\nabla}$ we deduce
\eq{ \label{fundamental1}
\overrightarrow{\nabla} \omega &= \nabla (dx^I \star \omega_I) = \nabla (dx^I) \star \omega_I |_{\ov \phi} + \ov{\cal R}(dx^I) \star \ov{\cal R}(\partial) \omega_I \\
&= \partial \omega - \Gamma \star \omega\,,
}
while for $\overleftarrow{\nabla}$ we find
\eq{ \label{fundamental2}
\overleftarrow{\nabla} \omega &= (dx^I \star \omega_I)\nabla = (dx^I) \ov{\cal R}(\nabla) \star \ov{\cal R}(\omega_I) + dx^I \star \partial \omega_I \\
&= \partial \omega - \omega \star \Gamma\,.
}
So far we defined a covariant derivative acting on covectors.
The same procedure can be repeated  for vectors without obstructions. We find
\eq{ \label{anomalousvector}
\Delta^\star_\xi  \; \partial V = -&\ov{\cal R}(V) \star \partial \partial \ov{\cal R}(\xi)\,.
}
Along the lines of \eqref{anomalousvariation},   the
anomalous transformation $\ov{\cal R}(V) \star \partial \partial
\ov{\cal R}(\xi) = \partial \partial \xi \star V$ can again be
compensated by $\overleftarrow{\nabla}$ and $\overrightarrow{\nabla}$
by only changing the overall sign  in front of $\Gamma$. 
This is also consistent with the more axiomatic viewpoint.
Therefore, we  again have two consistent covariant 
derivatives
\eq{
\overrightarrow{\nabla}  V &= \nabla (V) =  \partial V + \Gamma \star V \\
\text{and} \qquad \overleftarrow{\nabla}  V &= (V) \nabla = \partial V + V \star \Gamma\,.
}
In contrast to covectors, where $\overrightarrow{\nabla} \omega$ was
especially simple, now $\overleftarrow{\nabla} V$ becomes  simple. Since the
basis vectors of $V$ and $\Gamma$ are next to each other, the
contraction in the second term simply gives $V^I \star \Gamma_I$.
In addition, the left-linearity of vectors is similar to the left-linearity of
$\overleftarrow{\nabla}$. 

All this suggests that  expressions simplify if we use
$\overrightarrow{\nabla}$ for covectors and $\overleftarrow{\nabla}$ for
vectors. This convention is  compatible with the contraction if,
similar to $\partial$, the covariant derivative acts without 
an ${\cal R}$-matrix on products\footnote{Using the 
same convention for the covariant derivative on covectors and vectors,
compatibility with the contraction is also satisfied, but now explicit
${\cal R}$-matrices appear
\eq{
\overrightarrow{\nabla}( V \star \omega) &= \overrightarrow{\nabla} V \star \omega |_{\ov \phi} + \ov{\cal R}(V) \star (\ov{\cal R}(\overrightarrow{\nabla})\omega) |_{\phi^2} = \partial (V \star \omega)\,,\\
\overleftarrow{\nabla} (V \star \omega) &= V \star  \omega\overleftarrow{\nabla} |_{\phi} + V \ov{\cal R}(\overleftarrow{\nabla}) \star \ov{\cal R}(\omega) |_{\ov \phi^2}  = \partial (V \star \omega)\,.\nonumber
}
}
\eq{
  \overleftarrow{\nabla}V \star \omega |_{\ov \phi} +V \star
  \overrightarrow{\nabla} \omega |_{ \phi^2} &= \partial (V \star
  \omega) + (V \star \Gamma) \star \omega  |_{ \phi} - V \star (\Gamma
  \star \omega)  |_{ \phi^2} \\
&= \partial (V \star \omega)\,.
}

\subsubsection*{Directional covariant derivative}

At last we define the directional covariant derivative of a vector $Y=Y^I\star\partial_I$
simply by multiplication with the directional vector $X= X^I\star\partial_I$. We can
multiply $X$ either from the left or from the right onto
$\overleftarrow{\nabla}$ or $\overrightarrow{\nabla}$ and therefore
have in total four different conventions. Two choices place $X$ and $Y$
next to each other while the other two separate $X$ and $Y$ by a
$\Gamma$. To define a star-torsion we will soon see that we better place
$X$ and $Y$ together. 

To make the contraction in the second term as
easy as possible, we define the directional derivative along $X$ of a vector $Y$  as
\eq{
\label{directcovariant}
\nabla_X Y := \overleftarrow{\nabla}_{\overrightarrow{X}} Y:&= X \star \overleftarrow{\nabla} Y |_{\phi} = X \star \partial Y + (X \otimes_\star Y) \star \Gamma  \\
&= X \star \partial Y + \langle X \otimes_\star Y \;,\;\Gamma \rangle_\star 
}
where the associator was inserted for convenience. 
Spelling out the contraction between $X \otimes_\star Y$ and $\Gamma$ according to \eqref{basistotheright} reveals a correction term
\eq{ \label{covariantonvectors}
\nabla_X Y = X^I \star \partial_I Y^J \star \partial_J &+ (X^I \star Y^J) \star \Gamma_{IJ}{}^K \star \partial_K \\&- F^{MNJ} (X_M \star \partial_N Y^I) \star \Gamma_{IJ}{}^K \star \partial_K\,.
}
Recalling \eqref{Rbasistotheright} we find
\eq{ \label{Rofcov}
\nabla_{\ov{\cal R}(Y)} \ov{\cal R}(X) = \ov{\cal R}(Y)^I \star \partial_I &\ov{\cal R}(X)^J \star \partial_J + (X^J \star Y^I) \star \Gamma_{IJ}{}^K \star \partial_K \\
&\! - F^{MNI} (X_M \star \partial_N Y^J ) \star \Gamma_{IJ}{}^K \star \partial_K\,.
}
This is very useful when computing the torsion in the next paragraph.

As a comment, please note that, in the noncommutative though still associative framework 
of \cite{Aschieri:2005zs}, the convention $\nabla_X Y =
\overleftarrow{\nabla}_{\overrightarrow{X}} Y$ is used, as well\footnote{
In \cite{Aschieri:2005zs}  they demand
$\nabla_X (f \star Y) = \ov {\cal R}(f) \star \nabla_{\ov{\cal R}(X)} Y$,
which is only possible if $X$ and $Y$ are placed next to each other. 
Also in equation 5.4 of \cite{Aschieri:2005zs}, the left action
$\overleftarrow{\nabla}$ is used.}.

\subsection{Torsion}

We define the star-torsion two-form as usual as the antisymmetrized
covariant derivative $\overrightarrow \nabla^{\wedge_\star}$ of the frame $dx^I$. 
As we are in a holonomic frame we  find
\eq{
T^K =  \overrightarrow \nabla^{\wedge_\star}  dx^K = \Gamma^K = (dx^I \wedge_\star dx^J) \star \Gamma_{IJ}{}^K \,.
}
Setting this to zero means
\eq{ \label{torsionzero}
 \Gamma_{[IJ]}{}^K = 0 \,.
}
We now want to reproduce the same result  from an analogue of the
familiar definition $T =
\nabla_X Y - \nabla_Y X - [X,Y]$ with an appropriate insertion of
${\cal R}$ matrices. Note that here, for the directional covariant
derivative, we used the convention \eqref{directcovariant}.
We contract $T^K$ with vectors $X$ and $Y$ and apply \eqref{twoformonvectors} to turn the $\wedge_\star$ into an antisymmetrization of $X$ and $Y$ 
\eq{
T(X,Y):\!&=\langle X \otimes_\star Y \; , \; T^K \star \partial_K
\rangle_\star \\
&= \langle X \otimes_\star Y \; , \; dx^I \wedge_\star dx^J \star
\Gamma_{IJ}{}^K \star \partial_K \rangle_\star \\
&= \langle X \otimes_\star Y - \ov{\cal R}(Y) \otimes_\star \ov{\cal R}(X) \; , \; dx^I \otimes_\star dx^J \star \Gamma_{IJ}{}^K \star \partial_K \rangle_\star \,.
}
In the second line we identify the $\Gamma$ terms from the covariant derivative \eqref{directcovariant}. By adding and subtracting the missing terms $X \star \partial Y = X(Y)$ we can reproduce the torsion via
\eq{
T(X, Y) &= {\nabla}_{{X}} Y - {\nabla}_{{\ov {\cal R}(Y)}}  \ov{\cal R}(X) - [X,Y]_\star\,. 
}
At this point we need a convention where $X$ and $Y$ are
next to each other, as otherwise $X$ and $Y$ would be  separated
by  $\Gamma$. An explicit computation of the torsion by inserting \eqref{covariantonvectors} and \eqref{Rofcov} gives
\eq{
T(X,Y) = (X^I \star Y^J + F^{IMN} X_M \star \partial_N Y^J ) \star (\Gamma_{IJ} - \Gamma_{JI})\,.
}
The torsion tensor for basis vectors comes out as
\eq{
\langle T( \partial_I, \partial_J) , dx^K \rangle_\star = \Gamma_{IJ}{}^K - \Gamma_{JI}{}^K\,.
}

\subsection{Riemann and Ricci tensor}

In an analogous manner  we can proceed to derive a star-generalization of the
Riemann curvature.
The curvature two-form can be defined as  the exterior covariant
derivative of the connection $\Gamma= dx^K \otimes_\star dx^I \star
\Gamma_{KI}{}^L \star \partial_L$ which we consider
 as a matrix-valued one form $\Gamma_K{}^L :=dx^I \star \Gamma_{KI}{}^L$
\eq{ 
R_K{}^L  = {\nabla}^{\wedge_\star}\, \Gamma_K{}^L  &= d \Gamma_K{}^L  - \Gamma  _K{}^P \wedge_\star \Gamma_P{}^L \,.
}
We can contract the matrix indices with basis vectors and write 
\eq{
\label{curvato}
R = dx^K \star R_{K}{}^L \star \partial_L = d\Gamma - \Gamma \wedge_\star \Gamma  \, , 
}
where $d$ is meant to act on the one form part of $\Gamma$. Using 
\eqref{curvato}, tensoriality of $R$ can be readily checked using the anomalous
transformation $\Delta^\star_\xi \Gamma = \partial \partial \xi$, the
nilpotency of $d$ in $d \partial \xi = dd \xi =  0$ and the ${\cal
  R}$-antisymmetry \eqref{wedgeantisymmetric}  of $\wedge_\star$.

As the curvature $R$ contains three basis one-forms, we can contract it with three vectors 
\eq{ \label{Riemannform}
R(X, Y, Z) :&=  \langle  (X \otimes_\star Y) \otimes_\star Z \; , \; R \rangle_\star 
\\ &=  \langle (X \otimes_\star Y){}^\phi \;, \;  (Z^K {}^\phi \star R_K{}^L {}^\phi \star \partial_L ) \rangle_\star \,.
}
Similar to the torsion, one expects  this to  match the alternative definition
\eq{ \label{Riemannvector}
 R(X, Y, Z)  = -\Big(\big({\nabla}_{{X}}  \bullet {\nabla}_{{Y}} \big) Z  - \big( {\nabla}_{{\ov{\cal R}(Y)}}\bullet {\nabla}_{{\ov{\cal R}(X)}}\big)Z - {\nabla}_{{[X, Y]_\star }} Z\Big) \,.
}
The minus sign in this definition is just a convention needed  to match both definitions.
In order to evaluate \eqref{Riemannvector},
we need to clarify the meaning of the composition $\bullet$ for the directional covariant
derivatives. 

Having a closer look   at \eqref{directcovariant}, one realizes that
$\nabla_X Y$ consists actually of two consecutive operations: First the
action of $\overleftarrow{\nabla} = \partial + \Gamma$ from the
right and second the contraction with $X$ from the left,  denoted in
the following by $i_X$.
Next,  we apply an associator to bracket $X$ and $Y$ together.  
Following this prescription, we can write
\eq{
\nabla_X Y :&= \big(i_X \, (Y) \big) \, \overleftarrow{\nabla} =
i_X^\phi ( Y^\phi \overleftarrow{\nabla}^\phi) \\&=  i_X  \partial Y+   i_X^\phi (Y^\phi \star \Gamma^\phi ) = X \star \partial Y + (X \otimes_\star Y) \star \Gamma\,.
}
The composition $\bullet$ in \eqref{Riemannvector} must then be understood as
the composition of both the left and right acting operators
\eq{
({\nabla}_{{X}}  \bullet {\nabla}_{{Y}} ) Z &= \big[(i_X \bullet i_Y) ( Z) \big] \,  \, (\overleftarrow{\nabla}  \bullet \overleftarrow{\nabla})\,.
}
In appendix \ref{appandixriemann}, we show explicitly the equivalency 
of \eqref{Riemannform} and \eqref{Riemannvector}, when we evaluate
these four operations in the appropriate  order. 

To compute the components of the star-Riemann tensor, we need to shift all basis vectors in \eqref{Riemannform} into the middle with \eqref{switching3} and \eqref{hannover96}
\eq{ \label{components1}
R = dx^K \otimes_\star &dx^I \wedge_\star dx^J \\ & \star \big[ \partial_I \Gamma_{KJ}{}^L - \Gamma_{KI}{}^P \star \Gamma_{PJ}{}^L - F_I{}^{AB} \partial_A \Gamma_{KJ}{}^P \star \Gamma_{PB}{}^L \, \big] \star \partial_L
}
and
\begin{eqnarray}
\label{components2}
(X \otimes_\star Y) \otimes_\star Z \!\!\!&=&\!\!\! \Big[ (X^J \star Y^I) \star Z^K
\\[0.1cm]
&&- F^{ABJ} X_A^\phi \star \partial_B (Y^I{}^\phi \star Z^K{}^\phi) 
- F^{ABI} \star (X^J \star Y_A )\star \partial_B Z^K  \nonumber\\[0.1cm]
&&- F^{ABJ} F^{CDI} X_A^\phi \star \partial_B (Y_C^\phi \star \partial_D Z^K {}^\phi) \Big] 
 \star \partial_J \otimes_\star \partial_I
 \otimes_\star \partial_K\,. \nonumber
\end{eqnarray}
We  denoted the indices in such a way that $R(X, Y, Z)$ can be
directly  read off by star-multiplying the $[\dots ]$ brackets from
\eqref{components1} and \eqref{components2}. 
The components of the star-curvature thus contain a correction term proportional to the $R$-flux 
\eq{ \label{Riemanncomponents}
R_{IJK}{}^L :=& \langle R(\partial_I, \partial_J, \partial_K) , dx^L \rangle_\star \\ =& 2 \, \partial_{[\underline{I}}  \Gamma_{K \underline{J}]}{}^L - 2 \, \Gamma_{K[\underline{I}}{}^M \star \Gamma_{M\underline{J}]}{}^L - 2 \, F_{[\underline{I}}{}^{AB} \, \partial_A \Gamma_{K \underline{J}]}{}^M \star \Gamma_{MB}{}^L\,.
}
For a torsion-free connection, one can directly check that the first
Bianchi identity
\eq{
R_{IJK}{}^L+R_{KIJ}{}^L+R_{JKI}{}^L=0
}
is still satisfied.
The second Bianchi-identity receives a correction that, at leading
order in the flux $F$, is related to the associator of three connections
\eq{
\frac{1}{2} \nabla_{[\underline{I}} &R_{\underline{JK}]M}{}^N=\\
&\quad\ \Big[ \left(\Gamma_{[\underline{I}M}{}^A \star \Gamma_{\underline{J}A}{}^B\right)\star
  \Gamma_{\underline{K}]B}{}^N-  
   \Gamma_{[\underline{I}M}{}^A \star \left(\Gamma_{\underline{J}A}{}^B\star
  \Gamma_{\underline{K}]B}{}^N\right)\Big] + O(F)\,.
}
Notice that we need to use the convention $\overrightarrow{\nabla}$
for covectors  and $\overleftarrow{\nabla}$ for vectors to cancel the terms of the form $\sim \partial \Gamma \star \Gamma$ and $\sim \Gamma \star \partial \Gamma$.
As usual, the Ricci tensor is the trace of the Riemann tensor
\eq{ \label{Riccitensor}
\text{Ric} (Y, Z) :&= \langle R(\partial_I, Y, Z) , dx^I \rangle_\star \, . 
}


\section{Features of a star-metric}

In gravity the fundamental field is not a connection but a metric 
$G\in T^*M\otimes T^*M$
that allows to measure distances on the manifold.  Given a metric one
then defines the Levi-Civita  connection to be the torision-free
connection that warrants  a covariantly constant metric.
In this section we will see that the generalization  of this
procedure  to  the
nonassociative case appears to be less straightforward. Since so far we
did not find a fully satisfying resolution of the encountered
obstacles, this section should be  understood as a first approach to
this problem. In most parts of this section, we restrict our
considerations to star-tensors which depend only on configuration space.

Before we move on, let us recall that in differential geometry the metric is used in two 
ways. First it provides a scalar product between two vectors from the
tangent space, i.e. 
\eq{
\label{scalarprod}
                        (v,w)^g= g_{ij} v^i w^j\,.
}
Second it is considered to be a duality map ${\cal G}:TM\to T^*M$ that allows to
lower indices
\eq{
                               {\cal G}(v)_j= g_{ij} v^j\,.
}
The scalar product in \eqref{scalarprod} is then identical to
$(v,w)^{\cal G}=\langle v,
{\cal G}(w)\rangle$. Moreover, the inverse metric can be used to raise indices
and of course one has
\eq{
                   {\cal G}^{-1}\big({\cal G}(v)\big) =v\,,\quad\qquad
                                      g^{ij}\left(g_{jk} v^k\right)=v^i     \,.      
}     

\subsection{Metric}

We introduce a star-metric $G$ as an ${\cal R}$-symmetric element in
$T^*{\cal M}\otimes_\star T^*{\cal M}$,
i.e. it satisfies
\eq{
             G(X,Y)=G(\ov{\cal R}(Y) ,\ov{\cal R}(X)) \,
}
with
\eq{
G(X,Y):\!&=\langle X \otimes_\star Y \; , \; (dx^I \otimes_\star dx^J)
\star g_{IJ}\rangle_\star \,.
}
Recalling \eqref{basistotheright} and \eqref{Rbasistotheright} one obtains
\eq{
0&=G(X,Y) -G(\ov{\cal R}(Y),\ov{\cal R}(X))\\
&= (X^I \star Y^J + F^{IMN} X_M \star \partial_N Y^J ) \star (g_{IJ} - g_{JI})\,
}
so that  $g$ must be symmetric
in the  usual sense $g_{IJ} = g_{JI}$. Turning this around, every
symmetric tensor gives rise to a star-metric.

\subsubsection*{{\cal R}-symmetric scalar product}
The first definition of the scalar product is 
\eq{
( V, W)_\star^g := \langle (V \otimes_\star W), g \rangle_\star\,.
}
With \eqref{symmetry} the ${\cal R}$-symmetry of $g$ translates into the ${\cal R}$ symmetry between the vectors
\eq{
\langle (V \otimes_\star W), g \rangle_\star = \langle \ov{\cal R}(W) \otimes_\star \ov{\cal R}(V) , g \rangle_\star\, .
}
For the easiest example where $v = v \star \partial_i$ and $w = w \star \partial_i$, this scalar product is 
\eq{
\label{starscalarprod}
( v, w)_\star^g := (v^i \star w ^j ) \star g_{ij}\,.
}

\subsubsection*{The metric as a star-duality map}

Similar to the usual case, we can also interprete the metric $g$ as 
the duality map ${\cal G}:  T{\cal M}
\xrightarrow{g} T^*{\cal M}$ acting through ${\cal G}(W) = \langle W, g
\rangle_\star \in T^*{\cal M}$. 
Let us again only consider the easiest example $v = v \star \partial_i$ and $w = w \star \partial_i$. When we compute $\langle v, {\cal G}(w)\rangle_\star$
one finds
\eq{
    (v,w)^{\cal G}=v^i \star (w ^j \star g_{ij})
}
which is not the same as the star-scalar product
\eqref{starscalarprod}.
In fact the two are related by applying an associator
\eq{
                  ( v, w)^g_\star= (v,w)^{\cal G}\vert_\phi\,.
}
As a consequence of the appearing associator, $(v,w)^{\cal G}$
is not ${\cal R}$-symmetric.

A second deviation from the usual case appears when one considers
the inverse of the star-duality map
${\cal G}^{-1}:  T^*{\cal M} \xrightarrow{g^{-1\star}} T{\cal M}$, which should satisfy
${\cal G}^{-1}\big({\cal G}(v)\big)=v$ for all $v\in T{\cal M}$. In components this reads
\eq{
\label{starinversea}
                  \big(v^k \star g_{kj}\big)\star (g^{-1\star})^{ji}=v^i
}
and deviates from the usual case in the sense that, due
to nonassociativity,  an inverse 
satisfying 
\eq{
\label{starinverseb}
g_{ij}\star (g^{-1\star})^{jk}=\delta_i{}^k
} 
does not satisfy \eqref{starinversea}. In the noncommutative but
associative case, a construction of a general star-inverse in the
sense of \eqref{starinverseb} was provided in
\cite{Aschieri:2005yw,Aschieri:2015roa}. 

\subsection{The star-inverse}
\label{inversepossible}

It is clear that in order to proceed along the usual lines, one needs a star-inverse of the
metric. Recall that the inverse of the metric appears explicitly
in the Levi-Civita connection and in the definition of the
Ricci-scalar. In general it is unclear whether a solution to \eqref{starinverseb}
exists.
However, as emphasized in the introduction, from the
string theory viewpoint, it is actually only up  to linear order 
in $R^{ijk}$ that the star-product is really trustable.
Recall that when the nonassociative product was derived in
\cite{Blumenhagen:2011ph}, it was done for a flat metric with
a constant $R$-flux, which is only a solution of the string equations
of motion up to linear order in $R$.  
In this section, we therefore consider first the construction
of the inverse of a scalar and second the construction
of an inverse of the star-metric up to linear order in the $R$-flux.

\subsubsection*{Star inverse of a scalar}

Let us consider the  simpler question of constructing the
star-inverse $f^{-1\star}$  of a scalar $f$, which  has to satisfy
\eq{ \label{inversedef}
f^{-1\star} \star (f \star g) = g \,,\qquad \forall g\,.
}
We sort this equation according to the derivatives acting on $g$. At zeroth order in derivatives of $g$ the star-product between $f$ and $g$ becomes a usual multiplication
\eq{
f \star g = f g + \partial_I g \cdot \dots\,.
}
When carrying out  the remaining star-product in \eqref{inversedef},
since all derivatives act only on $f$, 
we find at zeroth order in derivatives of $g$
\eq{
f^{-1\star} \star (f \star g) = (f^{-1 \star } \star f) \cdot g + \partial_I g \dots\,.
}
Since this must be equal to $g$, we conclude that $f^{-1 \star }$ has to satisfy 
\eq{
f^{-1 \star} \star f = 1\,.
}
For general $g$ this is a contradiction to \eqref{inversedef} unless
the associator of $f^{-1 \star }$ and $f$ trivializes, i.e.  $\phi( f^{-1 \star }, f, \,\cdot\,)
= 1$. Of course, we do not expect that this is a generic
situation\footnote{Nonassociative algebras satisfying $\phi( f, f, \,\cdot\,)=\phi( \,\cdot\,,f, f)
= 1$ are called alternative. See \cite{Kupriyanov:2016joc} for a recent discussion in
the context of nonassociative star-products. As can be seen from
\eqref{hertha}, for general momentum our star-product is not alternative.}.
However, there exist certain scalars for which the
star-inverse can be identified. Consider e.g. the exponentials
$f(x)=\exp(i\vec q\,\vec x)$ from section \ref{fundamentalfunctions}. As one can easily show,
in this case the $\star$-inverse is simply $f^{-1\star}(x)=\exp(-i\vec q\,\vec
x)$. Indeed this scalar satisfies
\eq{
                   \phi(f^{-1\star},f,\,.\,)=1\,,\qquad 
                   f^{-1\star}\star f =f^{-1\star}\cdot f =1\,.
}
As a matter of fact one can show that, for a map $h\in C^\infty(M)$,
the star-inverse
of ${\mathfrak h}(\vec x)=h(\vec q\,\vec x)$ is ${\mathfrak h}^{-1\star}= 1/h(\vec q\,\vec x)$.

\subsubsection*{Star-inverse of the metric}

Let us now come back to the metric $g_{ij}(x)$ and from now on proceed in linear
order in the $R$-flux. 
At this order, we try to find solutions to 
\eq{
g_{ij} \star g_R^{\star -1 jk}  = \delta_i^k \, + \, {\cal O}(R^2)\,, \qquad
 g_L^{\star -1 ij}  \star g_{jk}  = \delta_i^k \, + \, {\cal O}(R^2) 
}
where we distinguished between  a right- and a left-inverse.
Remarkably, up to linear order, one can  explicitly solve these equations
\eq{
\label{inverselin}
g_R^{\star -1 ij} &= g^{ij} - \tfrac{i l_s^4}{6 \hbar} R^{abc} \,p_c\, \, g^{im}   \, \partial_a g_{mn}  \,\partial_b g^{nj} + \tfrac{l_s^4}{12} \, R^{abc}  \, \partial_a g^{im}  \, \partial_b g_{mn}  \,\partial_c g^{nj}\,,\\[0.1cm]
g_L^{\star -1 ij} &= g^{ij} - \tfrac{i l_s^4}{6 \hbar} R^{abc} \,
p_c\, \partial_a g^{im}   \, \partial_b g_{mn}  \, g^{nj} - \tfrac{l_s^4}{12} \, R^{abc}  \, \partial_a g^{im}  \, \partial_b g_{mn}  \,\partial_c g^{nj}\,.
}
For these star-inverse metrics we observe: 
\begin{itemize}
\item{They are not symmetric any longer.
For their symmetric parts one finds $g_{L/R}^{\star -1 (ij)}=g^{ij}$ and the antisymmetric
parts are given by the linear corrections in \eqref{inverselin}.}
\item{The star-inverses are momentum dependent, even if the original
star-metric was not.}
\item{Taking into account  \eqref{hertha}, one realizes  that the left- and
the right-inverse differ by an associator. 
}
\end{itemize}
The latter point  is explicitly reflected by expressing
\eqref{inverselin} as
\eq{ 
\label{leftandright}
g_R^{\star - 1 ij} &= 2 g^{ij} - g^{im} \star (g_{mn}\star g^{nj}) \,,\\[0.1cm]
g_L^{\star - 1 ij} &= 2 g^{ij} - (g^{im} \star g_{mn}) \star g^{nj}\,.
}
In this form the inverses  are  very  similar to the inverse metric on the
Moyal-Weyl-Plane in \cite{Aschieri:2005yw}. 
However, this inverse does not satisfy \eqref{starinversea}, as
\eq{  
\label{solveobstacle}
\big(v^k \star g_{kj}\big)\star (g_R^{-1\star})^{ji}=v^i+{l_s^4\over 6}
       R^{abc} \,\partial_a v^k\, \partial_b g_{kj}\,\partial_c g^{ji}
       +{\cal O}(R^2)\,,
}
where the second term is in general not vanishing as 
$\phi(g_{ij},  g^{jk},\,.\,)\ne 1$.
As a consequence,  the existence of these
star-inverse metrics does not allow us to solve equations involving
the metric.

\subsection{Comments on Levi-Civita connection}

In this final section we discuss the consequences of the previous
discussion on the  construction of a star-Levi-Civita connection.
The latter  is a torsion-free, metric compatible  connection, i.e $\nabla g=0$.
In this section, we do not restrict to the space-time components
of the star-metric but also consider the momentum components.

In this general case, the condition for the star-inverse of the metric
reads
\eq{ \label{inverseconcrete}
\delta &= g \star g^{\star -1} \\&= dx^I \otimes_\star dx^J \star g_{IJ} \star g^{\star -1}{}^{AB} \star \partial_A \otimes_\star \partial_B
\\ &= dx^I \star \Big(g_{IJ}\star (g^{\star -1}{})^{JB} + F_A{}{}^{MN}  \partial_M \big({g_{IN}} \star (g^{\star -1}{})^{AB}\big) \Big) \star \partial_B \, ,
}
where the additional term is there to compensate the
shift of the basis vector and no obstacle in finding an inverse.
To embed the space-time dynamics into the phase space we make the
ansatz\footnote{Considering  the metric instead of the vielbein as the fundamental field is
also motivated by string theory, where  the star-product 
appears between the vertex operators and therefore between the
fluctuations of the metric.}
\eq{ 
\label{gansatz}
g = dx^i \otimes_\star dx^j \star g_{ij}(x) + \tfrac{ dp_i}{i \hbar} \otimes_\star  \tfrac{ dp_i}{i \hbar} \star \eta^{ij}\,.
}
From \eqref{inverseconcrete} we see that we need a compensating term for the shift of basis vectors giving
\eq{ \label{ginverse}
g^{\star -1}_{R} = g^{\star - 1 ij}_{R} \star \partial_i &\otimes_\star \partial_j + \eta_{ij} \star i\hbar \tilde \partial_p^i \otimes_\star i\hbar \tilde \partial_p^j \\&- \tfrac{l_s^4}{6} R^{amn} g^{bi} \star \eta_{aj} \star \partial_m g_{bn} \star \partial_i \otimes_\star \tilde \partial_p^j \, .
}
Vanishing torsion implies $\Gamma_{[IJ]}{}^K=0$ so that, 
proceeding  in the usual way, we arrive at the relation
\eq{ \label{Levi-civita}
(dx^I \otimes_\star  &dx^J \otimes_\star dx^K)  \\ 
&\star \big[ \partial_I g_{JK} + \partial_J g_{IK} - \partial_K g_{IJ} \big]  =(dx^I \otimes_\star dx^J ) \star  2 \,\Gamma_{IJ}{}^L \star dx^K \star g_{LK} \, \\
&\hspace{4.9cm}\, = 2 \, \Gamma \star g\,,
}
that needs to solved for $\Gamma$.
However, due to the appearing associator in \eqref{solveobstacle},
this is {\it not}  solved by  $\Gamma=(\sum \partial g)\star g^{\star
  -1}_{R}$,
where $\sum \partial g = (\partial_I g_{JK} + \partial_J g_{IK} - \partial_K g_{IJ}) $.

We observe that for the Levi-Civita connection, the
obstruction arising from  \eqref{solveobstacle} could in principle
be cured by  reordering  the brackets in the Levi-Civita connection
by hand, i.e. by defining its  covariant derivative on a covector as 
\eq{
\nabla^{\rm LC} \omega = \partial\omega -\Big(\big({\textstyle \sum} \partial g\big) \, \circ \, g^{\star - 1}_L\Big) (\omega)=
\partial \omega -   \big({\textstyle \sum} \partial g\big) \star
(g_L^{\star - 1} \star \omega) \,.
}
Note that this is just an ad-hoc measure that is not consistent 
with the  $\bullet\,$-composition introduced in \eqref{composition},
which involved an extra application of the associator.
 Most importantly, as pointed out in \cite{Mylonas:2013jha}, 
  when using the $\circ$ composition, one is  in
 general not considering the connection $\Gamma$ as an
 independent object\footnote{If this would be possible, then
$(A\circ B)\star 1=A\star (B\star 1)=A*B$, so that $(A\circ B)=A\star
B$, implying $(A\star B)\star C=(A\circ B)\star C=A\star (B\star C)$.
This  is not satisfied in the  nonassociative case
\cite{Mylonas:2013jha}.}. 

Since this implies a major deviation from the structure
introduced  so far, the precise justification of such a 
definition of the connection
is beyond the scope of this paper. Nevertheless, we  would like to
finish our analysis with some comments about the remaining step
of defining a star-Einstein Hilbert action.

\subsubsection*{Final consideration on Einstein-Hilbert action}

For defining an Einstein-Hilbert action (up to linear order in $R$),
one needs a measure $\mu$ that should transform
as a star-scalar density under star-diffeomorphisms
\eq{
          \delta_\xi \mu = \xi^I \star\partial_I \mu +(\partial_I
          \xi^I)\star \mu\,.
}
In this case the Einstein-Hilbert action
\eq{
\label{actionEH}
    S&=\int d^d x\, d^d p\,  \Big(\mu  \star \text{Ric} \Big)
}
is star-diffeomorphism invariant as
\eq{
      \delta_\xi S=\int d^d x\,  d^d p \;\partial_I\Big( \xi^I\star (\mu\star
      \text{Ric})\Big) \,.
}
We make the usual choice $\mu=\sqrt{g}$ and, in the spirit of the
comment in section \ref{fundamentalfunctions}, consider it as
an elementary object that only depends on $x$.

Since we have the aforementioned bracketing issue in the Levi-Civita
connection, in the following we will make some general comments while being
agnostic about the it.
Now we consider the embedding \eqref{gansatz} and restrict the action
to configuration space via
\eq{
S &=  \int d^dx\,  d^dp  \,\, \sqrt{g} \star \text{Ric} \star \delta(p) \, . 
} 
Following the discussion in section \ref{sec_tri}, 
the $\delta(p)$ embeds the configuration space into the phase
space as the $p_0=0$ leaf\footnote{One could also carry out  the
  momentum integral 
 without the $\delta(p)$, as  $p$ enters only through the star-product
 and therefore linearly. 
Taking into account that 
$ \int_{\mathbb{R}^d} d^d p \,\,\, p^\mu = 0$
one is confined to the $p_0=0$ leaf, anyway.}.
Let us now analyze the linear terms in $R^{abc}$.
Two terms in the Ricci scalar  have  the formal structure
\eq{
\label{Riccione}
S_1=\int d^d x \, \sqrt{g} \star g^{-1} \star \partial \Gamma=
\int d^d x \, \sqrt{g} \star g^{-1} \star \partial (\partial g \star g^{-1} ) \,.
}
where we neither specify the bracketing nor the order and leave
it also open whether the left or the right inverse of the metric appears.

In the spirit of the comment from the introduction we want to know
whether there exists the possibility that the linear $R$-flux
correction is a total derivative(so that the nonassociativity does
not leave any trace  in the action).
There are two sources of linear terms in the $R$-flux:
\begin{enumerate}
\item{They
can appear from the star-product between the objects that
only depend on the coordinates $x$, i.e.
$\{g_{ij},\sqrt{g},(g_{L/R}^{*-1})^{(ij)}\}$.}
\item{The star-inverse 
has a linear correction $(g_{L/R}^{*-1})^{[ij]}$ that depends linearly
on $R^{abc}$ and also on the momentum coordinates $p$.}
\end{enumerate}
\noindent
Terms from category 1 are becoming total derivatives once  they
are bracketed in the nested way of eq.\eqref{triproda}.
Terms of category 2 can be trivially absent if
$(g_{L/R}^{*-1})^{[ij]}$ is coupled to symmetrized indices $(ij)$.
Moreover, we observe that 
\eq{
\int d^d x \, \big(g_L^{*-1}\big)^{[ij]} \star \psi_{ij}\big\vert_{p_0=0} &=
\int d^d x\,   R^{abc}\, \partial_a g^{im} \,\partial _b g_{mn}\, \partial_c
       ( g^{nj} \psi_{ij})\\
&=
\int d^d x\,   \partial_c \Big( R^{abc}\, \partial_a g^{im}\, \partial _b g_{mn} 
       \, g^{nj}\, \psi_{ij} \Big)
}
so that the linear $R$-correction for a left-placed $g_L^{*-1}$ gives
a total derivative. Similarly, for a right-placed $g_R^{*-1}$ one finds
\eq{
\int d^d x \; \psi_{ij} \star\big(g_R^{*-1}\big)^{[ij]}\big\vert_{p_0=0} &=
\int d^d x\,   \partial_c \Big( R^{abc}\, \psi_{ij}\,  g^{im}\, \partial _a g_{mn} 
       \,\partial_b   g^{nj}\Big)\,.
}
Since in \eqref{Riccione} there appear only two $g^{*-1}$ factors,
there exist an order/bracketing that only gives total derivatives at
linear order in $R^{abc}$.

The remaining terms in the Ricci-scalar  are of the schematic form
\eq{
\label{lastsweat}
S_2=\int d^d x \, \sqrt{g} \star g^{-1} \star \Gamma \star  \Gamma=
\int d^d x\,  \sqrt{g} \star  g^{-1} \star \partial g\star
g^{-1}  \star \partial g \star g^{-1}   \, .
}
Here we have three factors of $g^{*-1}$ so that one of them
cannot be placed entirely to the left or to the right. A more detailed
look at the index structure reveals that  
some of these terms are not trivially vanishing (i.e. 
coupling to a symmetrized pair of indices). 

Thus, we conclude that, irrespective of the bracketing/ordering,
there is no obvious reason why these corrections linear in $R^{abc}$
should give a total derivative. This could only happen via some
cancellations of terms, which however depends on the details
of the ordering/bracketing.
If linear effects remain, these will be of sixth order in derivatives
and are expected to break the usual diffeomorphism
symmetry.  This  makes it questionable whether  they have anything to
do with string theory.

From the string theory perspective, we recall from
\cite{Blumenhagen:2011ph} 
that the tri-product was derived for tachyon vertex operators only, while 
already the definition of a graviton vertex operator in a linear R-flux
background was not achieved in a straightforward manner.
Therefore, one might be sceptical about  the simple appearance  of the tri-product
between metric factors in the first place. In view of \cite{Blumenhagen:2011ph},
another possibility could be that the metric itself (and not only its
star-inverse) receives some order $R$-corrections.
Of course, all this is very speculative so that we stop here.

\section{Conclusions}

In this paper, in a step by step procedure  we have (re-)derived the salient 
structure of a nonassociative differential geometry that is based
on the nonassociative star-product arising for the closed
string moving in a constant nongeometric $R$-flux background.
Remarkably, even without associativity is was possible
to generalize the notions of diffeomorphisms, tensors, covariant derivatives,
torsion and curvature. This was possible, as mathematically
one is dealing with still  a special way of how associativity is broken,
namely that its information is encoded in an ${\cal R}$-matrix
and an associator $\phi$. Such a structure, namely the differential
geometry associated to a quasi-Hopf algebra, 
was recently 
developed from a very mathematical and abstract point of view in 
\cite{Barnes:2015aa,Barnes:2015bb,Barnes:2016cjm}.
In an attempt to make these results more accessible to physicists, 
we tried to motivate and clarify the appearing structure for
our concrete $R$-flux example from a  bottom up perspective.

As in \cite{Barnes:2015aa,Barnes:2015bb,Barnes:2016cjm}, the gravity theory could be well developed up to the
point where a metric and its Levi-Civita connection are introduced.
We argued that due to the nonassociativity, the star-metric
generically does not satisfy the usual relations for pulling up and down
indices. Up to linear order in the flux,  left/right-inverses of the
metric could be
identified that however were not symmetric and did not
allow a calculus, where equations could be solved.
Of course, it could well be that we are missing
a resolution of all these problems but it could also indicate that
there is something seriously wrong  about introducing a metric
on such spaces. 

At the end we were pointing out that maybe one needs
to define the action of the Levi-Civita connection in a different
way that employs the $\circ\,$-composition introduced in
\cite{Mylonas:2013jha}. 
Whether this major deviation from the structure
introduced  before leads to a consistent nonassociative gravity 
theory remains to be seen. 
Finally, we were commenting on the construction of 
a star-Einstein-Hilbert action and generally discussed 
whether it could be possible that, up to
first order in the $R$-flux, all effects of nonassociativity disappear
after restricting to the $p_0=0$ leaf. Of course this discussion
only becomes truly relevant after the issue about the 
definition of the star-Levi-Civita connection has been resolved.

Let us close by mentioning again that,  at the momentary state of
affairs,  it is an open question
whether such a  nonassociative gravity  theory based on the concept of  
star-diffeomorphisms has really anything to do with string or double
field theory, but it is certainly a viable and interesting possibility
that deserves further studies in mathematical physics.

\vskip2em
\noindent
\emph{Acknowledgments:} We are grateful to G. Barnes, P. Schupp and
R. Szabo for discussions.


\appendix

\section{Hopf algebra approach}
\label{app_hopf}

Above we derived an adjusted Leibniz rule to make the star-product behave covariantly. Mathematically this can be captured by Hopf algebras. We did not want to present our results in the abstract language of Hopf algebras for readability. Thus instead of being mathematically precise we will only give a short introduction into the topic to understand how the twisted Leibniz rule appears naturally in the context of Hopf algebras. Important for us are now only the multiplication $\mu$ and the coproduct $\Delta$ of the Hopf algebra. As usual the multiplication takes two objects and multiplies them to one. The comultiplication does the opposite. It takes one object and gives out two. Therefore
\eq{
\mu&: H \otimes H \rightarrow H \,,\\
\Delta&: H \rightarrow H\otimes H \, .
}
The Hopf algebra we are interested in is the universal enveloping algebra of the diffeomorphisms. It consists of what is usually denoted by $\delta_\xi$, thus the differential operator that becomes the actual transformation when acting on for instance a scalar $\delta_\xi \phi = \xi^\mu \partial_\mu \phi$. The multiplication is the usual one while the coproduct is
\eq{ \label{untwistedcoproduct}
\Delta (\delta_\xi) = \delta_\xi \otimes 1 + 1 \otimes \delta_\xi \, . 
}
Now it is clear that we should interpret the comultiplication $\Delta$ as the Leibniz rule for differentiation. For instance when acting on the product of two scalars we have
\eq{
\delta_\xi (\phi \psi) = (\delta_\xi \phi) \, \psi + \phi \, (\delta_\xi \psi) = \mu\big( \delta_\xi \phi \otimes \psi + \phi \otimes \delta_\xi \psi \big)= \mu \circ  \Delta(\delta_\xi) (\phi \otimes \psi) \, . 
} 
The left side can also be written as $\delta_\xi \, \mu(\phi \otimes \psi)$ which can be compared with the right side to 
\eq{
\delta_\xi  \mu = \mu \circ \Delta(\delta_\xi)\,.
}
We will now deform our product with a twist ${\cal F}$ as above \eqref{starproduct} to
\eq{
\mu_\star (f \otimes g):= \mu \circ {\cal F}^{-1} = f \star g
}
and demand the product to be compatible with the coproduct. A short calculation gives a twisted Leibniz rule
\eq{
\delta_\xi (\phi \star \psi) &= \delta_\xi \big( \mu \circ {\cal
  F}^{-1} (\phi \otimes \psi)\big) = \mu \circ \Delta(\delta_\xi) \circ {\cal F}^{-1} (\phi \otimes \psi)\\ &= \mu \circ {\cal F}^{-1} \circ {\cal F} \circ \Delta(\delta_\xi) \circ {\cal F}^{-1} (\phi \otimes \psi) \\
:&= \mu_\star \circ \Delta_\star(\delta_\xi) (\phi \otimes \psi)
}
where the new coproduct $\Delta_\star(\delta_\xi):= {\cal F}\Delta(\delta_\xi) {\cal F}^{-1}$ dictates the twisted Leibniz rule. As one can show these twisted objects still satisfy the axioms of a Hopf algebra or a generalization called quasi-Hopf algebra. So far we only used the generators $\partial_\mu$ for which we can calculate $\Delta_\star(\partial_\mu) = \Delta(\partial_\mu)$. In contrast to this we get 
\eq{ \label{momentumcoproduct}
\Delta_\star(\tilde \partial_p^\mu) = \Delta(\tilde \partial_p^\mu) + \tfrac{i l_s^4}{6 \hbar} R^{\mu \nu \rho} (\partial_\nu \otimes \partial_\rho)
}
and therefore the additional terms in the Leibniz rule as in \eqref{werder}.

\section{Computing the Riemann tensor}
 \label{appandixriemann}

 In this appendix we provide the  details on the evaluation  of
 \eq{ \label{fullriemann}
- R(X, Y, Z)  = ({\nabla}_{{X}}  \bullet {\nabla}_{{Y}} ) Z  - ( {\nabla}_{{\ov{\cal R}(Y)}}\bullet {\nabla}_{{\ov{\cal R}(X)}})Z - {\nabla}_{{[X, Y]_\star }} Z \,.
}
As discussed in the main text after \eqref{Riemannvector}, we need to interpret the $\bullet$ as a composition of left and right actions of the directional covariant derivative
\eq{
({\nabla}_{{X}}  \bullet {\nabla}_{{Y}} ) Z &= \big[(i_X \bullet i_Y) ( Z) \big] \,  \, (\overleftarrow{\nabla}  \bullet \overleftarrow{\nabla}) \,.
}
Recall that in $\nabla_X Y$ first $\overleftarrow{\nabla}$ is carried out
and afterwards $X$ acts as a contraction denoted by $i_X$. 
In addition, we have to respect the order of
$\nabla_X$ and $\nabla_Y$. Indicating the order by a subscript, we have 
altogether
\eq{
({\nabla}_{{X}}  \bullet {\nabla}_{{Y}} ) Z &= \big[(i_{X(4)} \bullet i_{Y(2)}) ( Z) \big] \,  \, (\overleftarrow{\nabla} _{(1)} \bullet \overleftarrow{\nabla}_{(3)})\,.
}
We apply the first covariant derivative by bringing $Z$ and
$\overleftarrow{\nabla}_{(1)}$ together. The scalar product between
$Z$ and the first matrix index of $\Gamma$ is carried out directly $Z
\star \Gamma = Z^M \star \Gamma_M$ followed by  bringing $i_Y$
together with $Z$.
Thus, the computation proceeds as
\eq{
\big[&(i_{X(4)} \bullet i_{Y(2)}) (Z) \big] \,  \, (\overleftarrow{\nabla} _{(1)} \bullet \overleftarrow{\nabla}_{(3)})\\
 &= (i_{X(4)} \bullet i_{Y(2)})^\phi  \,  \,  \big[ Z^\phi  \,  \,  (\overleftarrow{\nabla}{}_{(1)}  \bullet \overleftarrow{\nabla}{}_{(3)})^\phi \big]\\
&= (i_{X(4)} \bullet i_{Y(2)})^\phi  \,  \,   \big[ \partial Z^\phi   \,  \,  \overleftarrow{\nabla}{}_{(3)}^\phi \big] + (i_{X(4)} \bullet i_{Y(2)})^\phi  \,  \,  \big[ Z{}^M{}^\phi{} \star ( \Gamma_M {} \;\,\,  \overleftarrow{\nabla}{}_{(3)})^{ \phi} \big] \\
&= \big[i_{X(4)}^\phi ( Y^\phi  \star  \partial Z^\phi) \big]  \,  \,   \overleftarrow{\nabla}{}_{(3)}  + \big(\big[ i_{X(4)}^\phi ( Y^\phi  \star   Z{}^M{}^\phi) \big]^{\ov\phi}   \, \star   \,   \Gamma^{\ov\phi}_M \big) \,\,  \overleftarrow{\nabla}{}_{(3)}^{\ov\phi} \, .
}
Next, the second covariant derivative $\overleftarrow{\nabla}{}_{(3)}$ and afterwards $i_X$ are applied
\eq{ \label{riemannpart}
 = &
i_{X(4)}^{\phi\phi'}  \,  \,  \big[ ( Y^\phi  \star  \partial Z^\phi){}^{\phi'}   \overleftarrow{\nabla}_{(3)}^{\phi'} \big] + i_{X(4)}^{\phi\phi'} \big[( Y^\phi  \star   Z^K{}^\phi){}^{\phi'}   \star (\Gamma_K \;\; \overleftarrow{\nabla}_{(3)}){}^{\phi'} \big] \\
= &  X^\phi  \star \partial ( Y^\phi  \star  \partial Z{}^\phi) 
+ [X^\phi \star (Y^\phi \star \partial Z^K{}^\phi)
]  \star \Gamma_K   \\
&+ X{}^{\phi}{}^{\phi'} \star \partial \big[ (Y{}^{\phi} \star Z^K{}{}^{\phi}){}^{\phi'} \star \Gamma{}^{\phi'}_K \big] + \big[ (X \star Y)  \star   Z^K \big] \star   (\Gamma_K{}^{P} \star \Gamma_P){} \,.
}
In this formula we placed the brackets and the derivative $\partial$
in such a way that  they reflect,
which objects have to be contracted with each other. 
For instance in the first term of \eqref{riemannpart}, the derivative
is contracted with $X$. After applying the Leibniz rule for 
$\partial ( Y^\phi  \star  \partial Z{}^\phi)$, this contraction must
be kept in mind.

When computing the other terms in \eqref{fullriemann}, one realizes
that the first term in \eqref{riemannpart} is canceled partly by 
$( {\nabla}_{{\ov{\cal R}(Y)}}\bullet {\nabla}_{{\ov{\cal R}(X)}})Z$
and partly by $\nabla_{[X,Y]_\star} Z$. The other term
from $\nabla_{[X,Y]_\star} Z$ cancels the $X \star \partial Y \star  Z
\star \Gamma$ part in 
the third term of \eqref{riemannpart}. 
The remaining  two terms which have to cancel in \eqref{riemannpart}
arise  from the second and third term and are both of the form $X \star
Y \star \partial Z \star \Gamma$. In one term $\partial$ is contracted
with $Y$ and in the other $\partial$ is contracted with $X$. 
These terms cancel crosswise against  similar terms appearing
in $( {\nabla}_{{\ov{\cal R}(Y)}}\bullet {\nabla}_{{\ov{\cal R}(X)}})Z$.

After all these cancellations, the Riemann-tensor \eqref{fullriemann}
simplifies to
\eq{
\label{almostfinal}
- R(X, Y, Z) =& \big( (X \star Y) \star Z^M \big) \star \partial
\Gamma_M + \big(( X\star Y) \star Z^M \big) \star (\Gamma_M{}^P \star
\Gamma_P )\\
& - X \leftrightarrow^{\ov{\cal R}} Y\, .
}
Recalling the discussion after \eqref{riemannpart},  in the first term
of \eqref{almostfinal}, $X$ is
contracted with $\partial$. This is in contrast to
the rule that always a vector is contracted with the nearest
neighboring form\footnote{Notice that the contraction in the second term
  $\sim \Gamma \star \Gamma$ comes out correctly according to this
  rule.}. To bring this into the usual notation, we switch the first
term with its ${\cal R}$-permuted term and find with \eqref{basistotheright} and \eqref{Rbasistotheright}
\eq{
- R(X, Y, Z) =& - \big( (X \star Y) \star Z^M \big) \star \partial \Gamma_M + \big(( X\star Y) \star Z^M \big) \star (\Gamma_M{}^P \star \Gamma_P )\\& - X \leftrightarrow^{\ov{\cal R}} Y \,.
}
Now the notation matches the one in \eqref{Riemannform}, where the
vector $Y$ is contracted with the form $\partial$ (see also
\eqref{components1} and \eqref{components2}). By utilizing
\eqref{symmetry} to transfer the antisymmetrization on the vector side
towards the form 
side, we indeed find
\eq{
- R(X, Y, Z) &= \big( (X \star Y) \star Z^K\big) \star \big(- d\Gamma_K{} + \Gamma_K{}^P \wedge_\star \Gamma_P\big) \,.
}
This  matches the definition of the star-Riemann curvature  as the exterior covariant derivative of the connection $\Gamma$ in \eqref{Riemannform}.


\clearpage
\bibliography{references}  

\providecommand{\href}[2]{#2}\begingroup\raggedright\begin{thebibliography}{10}

\bibitem{Blumenhagen:2010hj}
R.~Blumenhagen and E.~Plauschinn, ``{Nonassociative Gravity in String
  Theory?},'' {\em J. Phys.} {\bf A44} (2011) 015401,
\href{http://www.arXiv.org/abs/1010.1263}{{\tt 1010.1263}}.

\bibitem{Lust:2010iy}
D.~L{\"u}st, ``{T-duality and closed string non-commutative (doubled)
  geometry},'' {\em JHEP} {\bf 12} (2010) 084,
\href{http://www.arXiv.org/abs/1010.1361}{{\tt 1010.1361}}.

\bibitem{Bouwknegt:2004ap}
P.~Bouwknegt, K.~Hannabuss, and V.~Mathai, ``{Nonassociative tori and
  applications to T-duality},'' {\em Commun. Math. Phys.} {\bf 264} (2006)
  41--69,
\href{http://www.arXiv.org/abs/hep-th/0412092}{{\tt hep-th/0412092}}.

\bibitem{Plauschinn:2012kd}
E.~Plauschinn, ``{Non-geometric fluxes and non-associative geometry},'' {\em
  PoS} {\bf CORFU2011} (2011) 061,
\href{http://www.arXiv.org/abs/1203.6203}{{\tt 1203.6203}}.

\bibitem{Blumenhagen:2014sba}
R.~Blumenhagen, ``{A Course on Noncommutative Geometry in String Theory},''
  {\em Fortsch. Phys.} {\bf 62} (2014) 709--726,
\href{http://www.arXiv.org/abs/1403.4805}{{\tt 1403.4805}}.

\bibitem{Blumenhagen:2011ph}
R.~Blumenhagen, A.~Deser, D.~L{\"u}st, E.~Plauschinn, and F.~Rennecke,
  ``{Non-geometric Fluxes, Asymmetric Strings and Nonassociative Geometry},''
  {\em J. Phys.} {\bf A44} (2011) 385401,
\href{http://www.arXiv.org/abs/1106.0316}{{\tt 1106.0316}}.

\bibitem{Condeescu:2012sp}
C.~Condeescu, I.~Florakis, and D.~L{\"u}st, ``{Asymmetric Orbifolds,
  Non-Geometric Fluxes and Non-Commutativity in Closed String Theory},'' {\em
  JHEP} {\bf 04} (2012) 121,
\href{http://www.arXiv.org/abs/1202.6366}{{\tt 1202.6366}}.

\bibitem{Chatzistavrakidis:2012qj}
A.~Chatzistavrakidis and L.~Jonke, ``{Matrix theory origins of non-geometric
  fluxes},'' {\em JHEP} {\bf 02} (2013) 040,
\href{http://www.arXiv.org/abs/1207.6412}{{\tt 1207.6412}}.

\bibitem{Andriot:2012vb}
D.~Andriot, M.~Larfors, D.~L{\"u}st, and P.~Patalong, ``{(Non-)commutative
  closed string on T-dual toroidal backgrounds},'' {\em JHEP} {\bf 06} (2013)
  021,
\href{http://www.arXiv.org/abs/1211.6437}{{\tt 1211.6437}}.

\bibitem{Condeescu:2013yma}
C.~Condeescu, I.~Florakis, C.~Kounnas, and D.~L{\"u}st, ``{Gauged
  supergravities and non-geometric Q/R-fluxes from asymmetric orbifold
  CFT`s},'' {\em JHEP} {\bf 10} (2013) 057,
\href{http://www.arXiv.org/abs/1307.0999}{{\tt 1307.0999}}.

\bibitem{Bakas:2015gia}
I.~Bakas and D.~L{\"u}st, ``{T-duality, Quotients and Currents for
  Non-Geometric Closed Strings},'' {\em Fortsch. Phys.} {\bf 63} (2015)
  543--570,
\href{http://www.arXiv.org/abs/1505.04004}{{\tt 1505.04004}}.

\bibitem{Siegel:1993xq}
W.~Siegel, ``{Two vierbein formalism for string inspired axionic gravity},''
  {\em Phys. Rev.} {\bf D47} (1993) 5453--5459,
\href{http://www.arXiv.org/abs/hep-th/9302036}{{\tt hep-th/9302036}}.

\bibitem{Siegel:1993th}
W.~Siegel, ``{Superspace duality in low-energy superstrings},'' {\em Phys.
  Rev.} {\bf D48} (1993) 2826--2837,
\href{http://www.arXiv.org/abs/hep-th/9305073}{{\tt hep-th/9305073}}.

\bibitem{Hull:2009mi}
C.~Hull and B.~Zwiebach, ``{Double Field Theory},'' {\em JHEP} {\bf 09} (2009)
  099,
\href{http://www.arXiv.org/abs/0904.4664}{{\tt 0904.4664}}.

\bibitem{Hohm:2010jy}
O.~Hohm, C.~Hull, and B.~Zwiebach, ``{Background independent action for double
  field theory},'' {\em JHEP} {\bf 07} (2010) 016,
\href{http://www.arXiv.org/abs/1003.5027}{{\tt 1003.5027}}.

\bibitem{Aldazabal:2013sca}
G.~Aldazabal, D.~Marques, and C.~Nunez, ``{Double Field Theory: A Pedagogical
  Review},'' {\em Class.Quant.Grav.} {\bf 30} (2013) 163001,
\href{http://www.arXiv.org/abs/1305.1907}{{\tt 1305.1907}}.

\bibitem{Berman:2013eva}
D.~S. Berman and D.~C. Thompson, ``{Duality Symmetric String and M-Theory},''
  {\em Phys.Rept.} {\bf 566} (2014) 1--60,
\href{http://www.arXiv.org/abs/1306.2643}{{\tt 1306.2643}}.

\bibitem{Hohm:2013bwa}
O.~Hohm, D.~L{\"u}st, and B.~Zwiebach, ``{The Spacetime of Double Field Theory:
  Review, Remarks, and Outlook},'' {\em Fortsch.Phys.} {\bf 61} (2013)
  926--966,
\href{http://www.arXiv.org/abs/1309.2977}{{\tt 1309.2977}}.

\bibitem{Blumenhagen:2013zpa}
R.~Blumenhagen, M.~Fuchs, F.~Ha{\ss}ler, D.~L{\"u}st, and R.~Sun,
  ``{Non-associative Deformations of Geometry in Double Field Theory},'' {\em
  JHEP} {\bf 04} (2014) 141,
\href{http://www.arXiv.org/abs/1312.0719}{{\tt 1312.0719}}.

\bibitem{Blumenhagen:2014gva}
R.~Blumenhagen, F.~Ha{\ss}ler, and D.~L{\"u}st, ``{Double Field Theory on Group
  Manifolds},'' {\em JHEP} {\bf 02} (2015) 001,
\href{http://www.arXiv.org/abs/1410.6374}{{\tt 1410.6374}}.

\bibitem{Blumenhagen:2015zma}
R.~Blumenhagen, P.~d. Bosque, F.~Ha{\ss}ler, and D.~L{\"u}st, ``{Generalized
  Metric Formulation of Double Field Theory on Group Manifolds},'' {\em JHEP}
  {\bf 08} (2015) 056,
\href{http://www.arXiv.org/abs/1502.02428}{{\tt 1502.02428}}.

\bibitem{Deser:2014wva}
A.~Deser, ``{Star products on graded manifolds and α′-corrections to Courant
  algebroids from string theory},'' {\em J. Math. Phys.} {\bf 56} (2015),
  no.~9, 092302,
\href{http://www.arXiv.org/abs/1412.5966}{{\tt 1412.5966}}.

\bibitem{Deser:2015okd}
A.~Deser, ``{Star products on graded manifolds and $\alpha'$-corrections to
  double field theory},'' in {\em {34th Workshop on Geometric Methods in
  Physics (XXXIV WGMP) Bialowieza, Poland, June 28-July 4, 2015}}.
\newblock 2015.
\newblock
\href{http://www.arXiv.org/abs/1511.03929}{{\tt 1511.03929}}.
\newblock

\bibitem{Mylonas:2012pg}
D.~Mylonas, P.~Schupp, and R.~J. Szabo, ``{Membrane Sigma-Models and
  Quantization of Non-Geometric Flux Backgrounds},'' {\em JHEP} {\bf 09} (2012)
  012,
\href{http://www.arXiv.org/abs/1207.0926}{{\tt 1207.0926}}.

\bibitem{Bakas:2013jwa}
I.~Bakas and D.~L{\"u}st, ``{3-Cocycles, Non-Associative Star-Products and the
  Magnetic Paradigm of R-Flux String Vacua},'' {\em JHEP} {\bf 01} (2014) 171,
\href{http://www.arXiv.org/abs/1309.3172}{{\tt 1309.3172}}.

\bibitem{Mylonas:2013jha}
D.~Mylonas, P.~Schupp, and R.~J. Szabo, ``{Non-Geometric Fluxes, Quasi-Hopf
  Twist Deformations and Nonassociative Quantum Mechanics},'' {\em J. Math.
  Phys.} {\bf 55} (2014) 122301,
\href{http://www.arXiv.org/abs/1312.1621}{{\tt 1312.1621}}.

\bibitem{Aschieri:2005yw}
P.~Aschieri, C.~Blohmann, M.~Dimitrijevic, F.~Meyer, P.~Schupp, and J.~Wess,
  ``{A Gravity theory on noncommutative spaces},'' {\em Class. Quant. Grav.}
  {\bf 22} (2005) 3511--3532,
\href{http://www.arXiv.org/abs/hep-th/0504183}{{\tt hep-th/0504183}}.

\bibitem{Aschieri:2005zs}
P.~Aschieri, M.~Dimitrijevic, F.~Meyer, and J.~Wess, ``{Noncommutative geometry
  and gravity},'' {\em Class. Quant. Grav.} {\bf 23} (2006) 1883--1912,
\href{http://www.arXiv.org/abs/hep-th/0510059}{{\tt hep-th/0510059}}.

\bibitem{AlvarezGaume:2006bn}
L.~Alvarez-Gaume, F.~Meyer, and M.~A. Vazquez-Mozo, ``{Comments on
  noncommutative gravity},'' {\em Nucl. Phys.} {\bf B753} (2006) 92--127,
\href{http://www.arXiv.org/abs/hep-th/0605113}{{\tt hep-th/0605113}}.

\bibitem{Barnes:2015aa}
G.~E. {Barnes}, A.~{Schenkel}, and R.~J. {Szabo}, ``{Nonassociative geometry in
  quasi-Hopf representation categories I: Bimodules and their internal
  homomorphisms},'' {\em Journal of Geometry and Physics} {\bf 89} (Mar., 2015)
  111--152, \href{http://www.arXiv.org/abs/1409.6331}{{\tt 1409.6331}}.

\bibitem{Barnes:2015bb}
G.~E. {Barnes}, A.~{Schenkel}, and R.~J. {Szabo}, ``{Nonassociative geometry in
  quasi-Hopf representation categories II: Connections and curvature},'' {\em
  ArXiv e-prints} (July, 2015) \href{http://www.arXiv.org/abs/1507.02792}{{\tt
  1507.02792}}.

\bibitem{Barnes:2016cjm}
G.~E. Barnes, A.~Schenkel, and R.~J. Szabo, ``{Working with Nonassociative
  Geometry and Field Theory},'' \href{http://www.arXiv.org/abs/1601.07353}{{\tt
  1601.07353}}.
[PoSCORFU2015,081(2015)].

\bibitem{Aschieri:2015roa}
P.~Aschieri and R.~J. Szabo, ``{Triproducts, nonassociative star products and
  geometry of R-flux string compactifications},'' {\em J. Phys. Conf. Ser.}
  {\bf 634} (2015), no.~1, 012004,
\href{http://www.arXiv.org/abs/1504.03915}{{\tt 1504.03915}}.

\bibitem{Kupriyanov:2016joc}
V.~G. Kupriyanov, ``{Alternative multiplications and non-associativity in
  physics},'' in {\em {15th Hellenic School and Workshops on Elementary
  Particle Physics and Gravity (CORFU2015) Corfu, Greece, September 1-26,
  2015}}.
\newblock 2016.
\newblock
\href{http://www.arXiv.org/abs/1603.00218}{{\tt 1603.00218}}.
\newblock

\end{thebibliography}\endgroup
\bibliographystyle{utphys}


\end{document}